\documentclass[]{emulateapj}

\def\bibfiles{biblio}
\def\aareferences{\bibliographystyle{apj}
                  \bibliography{aajour,\bibfiles}}

\def\rmit#1{{\it #1}}              

\def\ie{\rmit{i.e.}}
\def\eg{\rmit{e.g.}}

\usepackage{natbib,graphicx}

\usepackage{color}

\usepackage{morefloats}

\shorttitle{Multiple scattering of $f-$modes by flux tubes}

\shortauthors{Felipe et al.}

\begin{document}

\title{Numerical simulations of multiple scattering of the $f-$mode by flux tubes}

\author{T. Felipe\altaffilmark{1}, A. Crouch\altaffilmark{1}, A. Birch\altaffilmark{2}}
\email{tobias@nwra.com}

\altaffiltext{1}{NorthWest Research Associates, Colorado Research Associates, Boulder, CO 80301, USA}
\altaffiltext{2}{Max-Planck-Institut f\"{u}r Sonnensystemforschung, Max-Planck-Str. 2, 37191 Katlenburg-Lindau, Germany}

\begin{abstract}
We use numerial simulations to study the absorption and phase shift of surface-gravity waves caused by groups of magnetic flux tubes. The dependence of the scattering coefficients with the distance between the tubes and their positions is analyzed for several cases with two or three flux tubes embedded in a quiet Sun atmosphere. The results are compared with those obtained neglecting completely or partially multiple scattering effects. We show that multiple scattering has a significant impact on the absorption measurements and tends to reduce the phase shift. We also consider more general cases of ensembles of randomly distributed flux tubes, and we have evaluated the effects on the scattering measurements of changing the number of tubes included in the bundle and the average distance between flux tubes. We find that for the longest wavelength incoming waves multiple scattering enhances the absorption, and its efficiency increases with the number of flux tubes and the reduction of the distance between them.

\end{abstract}



\section{Introduction}
\label{sect:introduction}

The study of solar oscillations has been broadly used to infer the properties of the Sun, giving birth to helioseismology \citep[see, \eg,][]{Christensen-Dalsgaard2002}. In recent years, there has been an increasing interest in using these techniques to reveal the subsurface nature of magnetic structures. A detailed theoretical description of wave interactions with magnetic fields is necessary to understand observations. When acoustic waves encounter magnetic flux tubes they excite tube waves, such as sausage and kink modes \citep{Spruit1981}. These tube waves propagate upward and downward and extract energy from the $p-$modes \citep{Bogdan+etal1996, Hindman+Jain2008, Jain+etal2009}. When the magnetic structure is stationary in comparison to the wave period, the outgoing wave is scattered at the same frequency as the incident wave. The scattering process produces outgoing propagating modes that transport energy away from the scatterer, and a continuous spectrum of horizontally evanescent wave modes which propagate vertically, the so-called jacket modes \citep{Bogdan+Cally1995}. Jacket modes appear surrounding the magnetic element, and are necessary to obtain continuity between the solution in the scattering magnetic structure and the adjacent quiet Sun. From the analytical point of view, the dipole scattering produced by isolated thin flux tubes has been studied for an incident $f-$mode \citep{Hanasoge+etal2008}, as well as the monopole component of both $f$ and $p$ modes \citep{Hindman+Jain2012}. Recently \citet{Felipe+etal2012a} analyzed the scattering of the $f-$mode by flux tubes using numerical simulations. They compared their numerical results for the phase shift with observational phase shifts obtained for the ensemble average of about 3400 small elements from MDI data studied by \citet{Duvall+etal2006}. They found that the observed dependence of the phase shift with wavenumber can be matched reasonably well with a simple flux tube model, while the observed variation with azimuthal order $m$ of the phase-shifts appears to depend on details of the ensemble averaging.

The has been important works on multiple scattering produced by several flux tubes over the last decades. \citet{Bogdan+Fox1991} studied the interaction of an acoustic plane wave with a pair of magnetic flux tubes. They distinguish three different scattering regimes, depending on the degree of coupling between the two flux tubes: an incoherent scattering regime, for large flux tube separations, where the scattering of each tube is independent of the presence of the companion tube; a coherent scattering regime, where the far field is affected by coherent scattering of the scattered waves from the two flux tubes; and a multiple scattering regime for sufficiently small flux tube separation. In the latter regime, the acoustic field scattered by one of the tubes deviates significantly from what would be expected in the absence of its companion flux tube. \citet{Keppens+etal1994} followed the formalism from \citet{Bogdan+Fox1991}, but considered tubes with a transition layer between the magnetic and non-magnetic atmosphere, allowing them to study resonant absorption. They examined the absorption of sound waves by bundles of magnetic flux tubes, containing from 2 to 19 individual tubes. They found that resonant absorption can be greatly enhanced by ensembles of flux tubes. Later, \citet{Tirry2000} found that the resonant absorption observed by \citet{Keppens+etal1994} is destroyed when the bottom boundary is opened and wave energy can leak out of the cavity through the magnetic flux tube. However, all of these studies neglected gravitational stratification. \citet{Hanasoge+Cally2009} investigated the scattering of acoustic waves by a pair of flux tubes in stratified media for the azimuthal orders $m=\pm 1$. They found that multiple scattering by two tubes at small separations can dramatically change the scattering properties. This work was extended to include axisymmetric ($m=0$) scattering by \citet{Hanson+Cally} (in preparation).

Interest in the study of the scattering by a bundle of magnetic flux tubes goes beyond the simple understanding of the physics involved in this phenomenon, since it is expected to play a key role in the response of some solar magnetic features to the wave field, like plage. \citet{Braun1995} observed that although the absorption by plage is only around two times smaller than in a sunspot, it produces no measurable phase shifts. The results from \citet{Hanasoge+Cally2009} point out that pairs of tubes can absorb effectively. According to their Figure 4, the scattered wave is phase-shifted, but the magnitude of this shift is very sensitive to the separation between the tubes. \citet{Jain+etal2009} modeled a plage region as a collection of non-interacting tubes in a stratified atmosphere. However, the theory of single scattering cannot properly describe the interaction of waves with a group of tubes \citep{Hanasoge+Cally2009}, which limits the relevance of their results. To date, no study has included a sufficiently large number of tubes to reliably estimate the expected absorption and phase shift measured in plages and account for the multiple scattering produced in the bundle of tubes. The investigation of other solar magnetic features like sunspots will also benefit from a better understanding of the scattering produced by groups of tubes, since the analysis of the acoustic field around sunspots has been proposed as a method to discern between the monolithic flux-tube and spaghetti sunspot models \citep{Bogdan+Fox1991}. Several papers \citep[\eg][]{Ryutova+etal1991, Labonte+Ryutova1993,Ryutova+Priest1993a,Ryutova+Priest1993b} have evaluated the absorption mechanism produced in sunspot models assembled from many individual magnetic flux tubes with random physical properties.

The aim of this work is to study the scattering properties of an ensemble of flux tubes by means of numerical simulations. We focus on the scattering produced by an incident $f$ mode, since its coupling with the flux tube is stronger than for the $p$ modes \citep{Bogdan+etal1996, Hanasoge+Cally2009}. A similar problem was recently studied by \citet{Daiffallah2013}. In Section \ref{sect:procedures} we describe the numerical procedures, including the numerical code, the background atmospheres, and the Hankel-Fourier analysis tools used in this work. Section \ref{sect:results} discusses the results obtained for the different numerical simulations, while the conclusions are presented in Section \ref{sect:conclusions}.

\section{Numerical procedures}
\label{sect:procedures}
We used the numerical code Mancha \citep{Khomenko+Collados2006, Felipe+etal2010a}. It numerically solves the nonlinear three-dimensional MHD equations for perturbations, that is, the equilibrium state is explicitly removed from the equations. The code computes spatial derivatives using fourth-order centered differences, and the solution is advanced in time using a fourth-order Runge-Kutta scheme. A perfectly matched layer (PML) boundary condition \citep{Berenger1994} is applied to the top and bottom boundaries in order to avoid wave reflections.

\begin{figure}[!ht] 
 \centering
 \includegraphics[width=9cm]{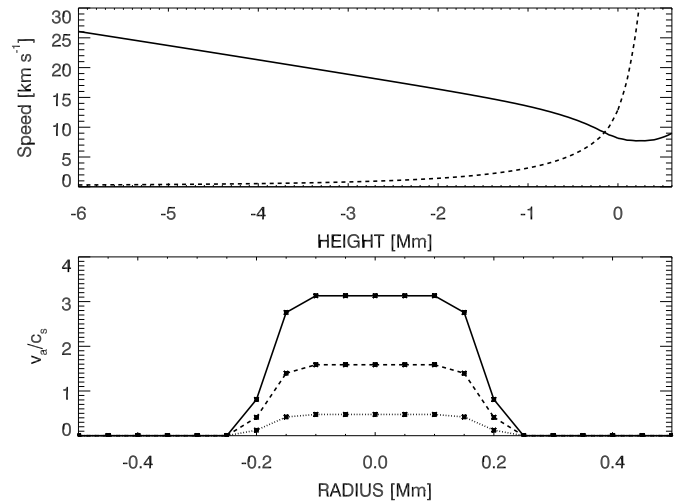}
  \caption{Characteristic speeds of the flux tube model. Top panel: variation of the sound speed ($c_S$, solid line) and Alfv\'en speed ($v_A$, dashed line) with height at the center of the flux tube. Bottom panel: radial variation of the ratio $v_a/c_s$ at $z=-0.5$ (dotted line), $z=0$ (dashed line), and $z=0.2$ Mm (solid line). The grid spacing is 50 km in the vertical and horizontal directions. }
  \label{fig:tube}
\end{figure}

We performed several simulations with different numbers and distributions of flux tubes. The method used to construct the tubes allows different size and field strength, but for simplicity, all the individual tubes used in this study have the same properties. They are constructed according to \citet{Pneuman+etal1986}, using the routines developed by \citet{Khomenko+etal2008a}. They have slight horizontal variations of the magnetic field and thermodynamic variables for radial distances below 100 km, with a maximum magnetic field around 1600 G. Over the next 100 km the magnetic field smoothly decreases until vanishing to avoid numerical problems due to a discontinuity. The tubes are embedded in a quiet Sun model S atmosphere \citep{Christensen-Dalsgaard+etal1996}, which was stabilized to avoid convection following \citet{Parchevsky+Kosovichev2007}. Figure \ref{fig:tube} shows the sound speed ($c_s$) and the Alfv\'en speed ($v_a$) of the flux tube model.

We are interested in studying the interaction of an $f$ mode with a group of flux tubes. According to \citet{Cameron+etal2008}, the vertical velocity of an $f$ mode wave packet which propagates in the $+x$ direction in a horizontally homogeneous atmosphere is given by:

\begin{equation}
\label{eq:fmode}
v_z(x,y,z,t)=Re\sum_k A_k e^{kz}e^{ik(x-x_0)-i\omega_kt}
\end{equation}

\noindent where $x$ and $y$ are the horizontal coordinates, $z$ is height, $Re$ means real part, $A_k$ are complex amplitudes, $x_0$ indicates the initial position of the wave packet, $t$ is time, $\omega_k=\sqrt{g_0k}$ is the eigenfrequency at horizontal wavenumber $k$, and $g_0$ is the gravitational acceleration. The initial conditions for the perturbations in velocity, pressure, and density that define this wave packet can be found in \citet{Felipe+etal2012a}. The complex amplitudes $A_k$ depend on the wavenumber. As an initial distribution, we have chosen a Gaussian centered at angular degree $L=kR_{\odot}=1000$, where $R_{\odot}$ is the solar radius, and with a half width of 600 in angular degree.

\begin{figure}[!ht] 
 \centering
 \includegraphics[width=7cm]{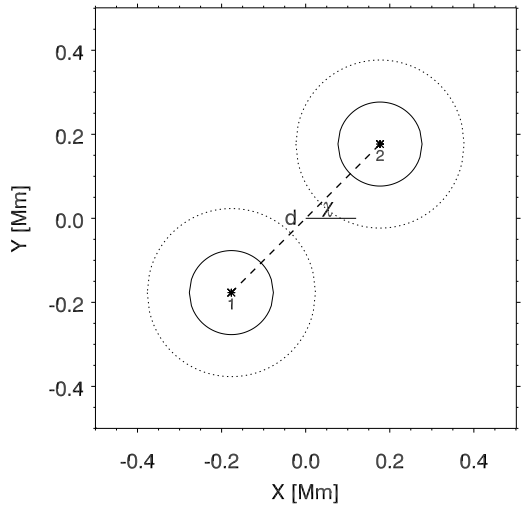}
  \caption{Schematic representation of the configuration of two tubes and the parameters that characterize it. The tubes 1 and 2 are separated by a distance $d$, with the line linking their centers inclined at angle $\chi$ to the $x$ axis. The solid circles contain the region of the tube where the magnetic field is around 1600 G, for larger radius the field strength decreases until it vanishes at the dotted circles.}
  \label{fig:tubos_pos}
\end{figure}

As a first step, we have performed several simulations including a pair of tubes in the atmosphere. The location of the tubes is characterized by the distance between the centers of the tubes $d=\|{\bf r}_1-{\bf r}_2\|$, where ${\bf r}_i$ is the horizontal position of the tube $i$ and $\|{\bf a}\|$ indicates the length of vector ${\bf a}$, and the angle of incidence $\chi$ corresponds to the angle between the horizontal direction of propagation of the $f$ mode and the separation vector that joins the center of the tubes ${\bf r}_1-{\bf r}_2$. Since in all cases the $f$ mode propagates in the $+x$ direction, $\chi$ is given by the angle between the $+x$ axis and ${\bf d}={\bf r}_1-{\bf r}_2$. Figure \ref{fig:tubos_pos} shows the configuration of the atmosphere with both tubes. In these simulations the computational domain spans from $z=-6$ Mm to $z=0.6$ Mm, where the photospheric level $z=0$ Mm is located at the height where the optical depth at 500 nm is equal to unity in the quiet Sun atmosphere. In the horizontal directions the domain extent is $x \in [-32.1,22.1]$ Mm and $y \in [-22,22]$ Mm, with the position $x=y=0$ corresponding to the middle of the vector $\bf{d}$. The spatial step is $50$ km in all directions. The initial position of the $f-$mode wave packet was set at $x_0=-27.1$ Mm, that is, out of the annular region which is going to be used for the Hankel analysis. The simulations are long enough to allow the $f$ mode to travel through the entire computational domain in the $x$ direction, corresponding to a duration of $T=135$ min. A summary of the simulations performed and their properties is shown in table \ref{tab:simulations}, including the number of tubes in the simulation and their distribution, the sections where they are discussed, and the corresponding figures.

\begin{table}[t]
\begin{center}
\caption[]{\label{tab:simulations}
          { Summary of the simulation runs}}
\begin{tabular*}{8cm}{@{\extracolsep{\fill}}ccccc}

\hline Section  & \# tubes & d [Mm] & $\chi$ [degrees] & Figures
\\
 \hline
 \ref{sect:d500a0}   & 1      &  -   &  -                &  \ref{fig:1tube},\ref{fig:absorption_maps},\ref{fig:alpha_L_x1000} \\
 \ref{sect:d500a0}   & 2      &  0.5 &  0                &  \ref{fig:alpha_d500a0},\ref{fig:phase_d500a0},\ref{fig:sigma_a},\ref{fig:phase_ntubes} \\
 \ref{sect:d500a90}  & 2      &  0.5 &  90               &  \ref{fig:alpha_d500a90},\ref{fig:phase_d500a90},\ref{fig:sigma_a}  \\
 \ref{sect:d500a45}  & 2      &  0.5 &  45               &  \ref{fig:alpha_d500a45},\ref{fig:phase_d500a45},\ref{fig:sigma_a} \\
 \ref{sect:d1000a0}  & 2      &  1   &  0                &  \ref{fig:alpha_d1000a0},\ref{fig:phase_d1000a0},\ref{fig:sigma_a},\ref{fig:phase_ntubes} \\
 \ref{sect:d4000a0}  & 2      &  4   &  0                &  \ref{fig:alpha_d4000a0},\ref{fig:phase_d4000a0},\ref{fig:phase_ntubes} \\
 \ref{sect:3tubes}   & 3      &  0.5 &  0                &  \ref{fig:alpha_3tubes},\ref{fig:phase_3tubes},\ref{fig:phase_ntubes} \\
 \ref{sect:14random} & 7      &  2.14 (av)  & random        &   \ref{fig:sigma_ensemble},\ref{fig:alpha_phase_ntubes},\ref{fig:phase_ntubes} \\ 
 \ref{sect:14random} & 14     &  1.07 (av)  & random        &  \ref{fig:alpha_phase_dij},\ref{fig:sigma_ensemble},\ref{fig:mixing_14tubes}\\ 
 \ref{sect:14random} & 14     &  2.14 (av)  & random        &  \ref{fig:alpha_phase_dij},\ref{fig:sigma_ensemble},\ref{fig:alpha_phase_ntubes},\ref{fig:phase_ntubes} \\ 
 \ref{sect:14random} & 14     &  3.20 (av)  & random        &  \ref{fig:alpha_phase_dij},\ref{fig:sigma_ensemble} \\ 
 \ref{sect:14random} & 28     &  2.14 (av)  & random        &  \ref{fig:sigma_ensemble},\ref{fig:alpha_phase_ntubes},\ref{fig:phase_ntubes} \\ 

\hline

\end{tabular*}
\end{center}
\end{table}

We have computed an equivalent 2D quiet Sun simulation, without the tubes being present, but with the same set up of the 3D cases. This simulation is used as a reference to compare its absorption and phase shift with those obtained for the simulations with the tubes, and to apply the appropriate procedures in order to retrieve the proper scattering properties caused by the magnetic elements \citep{Felipe+etal2012a}.

\begin{figure*}[!ht] 
 \centering
 \includegraphics[width=18cm]{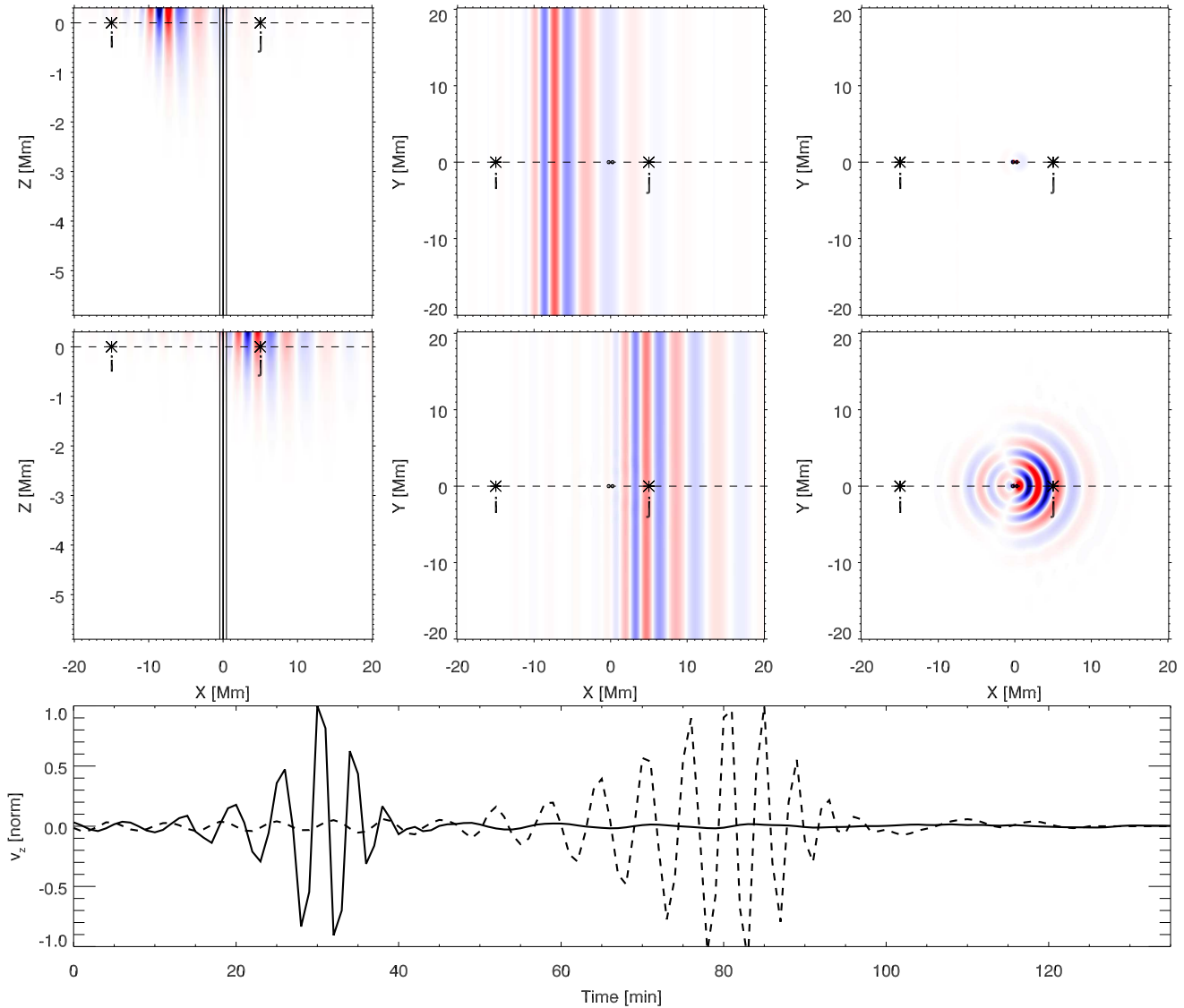}
  \caption{Wavefield in the simulation with two flux tubes with $d=0.5$ Mm and $\chi=0^o$. The top panels correspond to $t=50$ min and the middle panels to $t=80$ min. The top two rows show vertical velocity in the $xz$ plane with $y=0$ (left column), in the $xy$ plane with $z=0$ (middle column), and the scattered wave in the same $y$ plane (right column). The blue color corresponds to positive velocities (upflows) and the red color to negative velocities (downflows). Note that in the right column the color scale is 60 times more saturated than in the other two columns. The small circles around $x=y=0$ in the $xy$ plots and the vertical lines in the $xz$ plots correspond to the locations of the flux tubes. The bottom panel shows the time evolution of the vertical velocity at spatial points $i$ (solid line) and $j$ (dashed line). }
  \label{fig:wavefield}
\end{figure*}

Figure \ref{fig:wavefield} shows some snapshots of the wavefield produced by the simulation with two tubes aligned with the direction of propagation of the incident $f$ mode and separated by a distance $d=0.5$ Mm. The top panels show the wavefield just before the $f$ mode wave packet reaches the pair of tubes, while the middle panels correspond to the time step when the whole wave packet has passed though the flux tubes. The scattered wavefield (right column) is obtained by subtracting the 2D quiet Sun simulation from all the $xz$ planes in the 3D computation with the flux tubes embedded in the atmosphere. It is composed of outgoing propagating modes and jacket modes, although the latter are not visible in the plot. For a representation of the tube waves (kink and sausage modes) and jacket modes around a similar flux tube, see Figure 4 from \citet{Felipe+etal2012a}.

The simulations are analyzed by means of the Fourier-Hankel spectral decomposition method, discussed in \citet{Braun1995}. The vertical velocity is decomposed into ingoing and outgoing waves in an annular region, which allows us to detect the effect of the magnetic features on the wave field by analyzing the radially propagating Hankel components. In a spherical polar coordinate system $(\theta, \phi )$ the wave components take the form

\begin{eqnarray}
\label{eq:decomp}
\lefteqn{\Psi _m(\theta ,\phi,t)=e^{i(m\phi +2\pi \nu t)}\times} \nonumber\\
&&\times[A_m(L,\nu)H_m^{(1)}(L\theta )+B_m(L,\nu )H_m^{(2)}(L\theta )],
\end{eqnarray}

\noindent where $m$ is the polar azimuthal order, $H_m^{(1)}$ and $H_m^{(2)}$ are Hankel functions of the first and second kind, respectively, $L\equiv [l(l+1)]^{1/2}$ where $l$ is the spherical harmonic degree of the mode, $\nu$ is the temporal frequency, and $A_m$ and $B_m$ are the complex amplitudes of incoming and outgoing waves, respectively. 
 
The power of the ingoing and outgoing Hankel components has been averaged across the width of the $f-$mode ridge. At each given $L$, the power for the ingoing wave is determined as

\begin{equation}
\label{eq:Paverage}
|A_m(L)|^2=\int_{\nu_0(L)-\delta\nu}^{\nu_0(L)+\delta\nu}|A_m(L,\nu)|^2d\nu,
\end{equation}

\noindent where $\nu_0(L)$ is the $f-$mode frequency at degree $L$ and $\delta\nu\approx0.5$ mHz determines the width of the ridge. The same average is applied to the outgoing $B_m(L,\nu)$ components. The absorption coefficient $\alpha_m(L)$ along the ridge of the $f-$mode is then obtained as

\begin{equation}
\label{eq:alpha}
\alpha_m(L)=1-|B_m(L)|^2/|A_m(L)|^2.
\end{equation}
  
Some corrections are applied to the absorption coefficient in order to account for the numerical damping produced by the numerical diffusivity and the interaction of the $f-$mode with the top boundary, see \citet{Felipe+etal2012a} for details. The phase shift is given by

\begin{equation}
{\delta}(L) = \arg \left( \int_{\nu_0(L) - \delta \nu}^{\nu_0(L) + \delta \nu} 
 B_m (L, \nu) A^{*}_m (L, \nu) d\nu \right).
\label{eq:phase}
\end{equation}

\noindent The integral covers the same frequency bins from Equation \ref{eq:Paverage}. Finally, the phase shift of the quiet Sun reference simulation has been subtracted to the phase shift of the simulation with the flux tubes being present.

\section{Effects of multiple scattering on the absorption and phase shift}
\label{sect:results}

The resultant coefficients $A_m (L, \nu)$ and $B_m (L, \nu)$ for the ingoing and outgoing waves, respectively, of the Fourier-Hankel analysis were determined for waves within an annular domain delimited by the radial distances $R_{\rm min}=4.1$ Mm and $R_{\rm max}=20$ Mm, which provides a resolution in spherical degree $\Delta L=282.14$. The sampling in frequency is given by the duration of the simulations as $\Delta \nu = 1/T = 0.1235$ mHz. In the simulations with two flux tubes the center of the annular region was located at the center of the tube that first encounters the incident $f-$mode, that is, at the center of tube 1 in Figure \ref{fig:tubos_pos}.

We are interested in measuring the multiple scattering effects produced by the presence of another tube in the surroundings of a flux tube, and evaluating if the assumption of non-interacting flux tubes is a good approximation to describe the interaction of the $f-$mode with a bundle of magnetic elements. In order to quantify the impact of multiple scattering, we have compared the absorption coefficient of the simulation with two flux tubes with that produced by two isolated tubes located at the corresponding positions. The latter was obtained by performing two independent simulations, each of them with only one of the tubes being present. In the following, we will refer to the wave field of these simulations as {\bf u}$_i$, where $i$ indicates each of the tubes. The scattered wave field is obtained as  
 \begin{equation}
{\bf u}_{i}^{\rm sc}={\bf u}_{i}-{\bf u}_{\rm QS}
\label{eq:scattered}
\end{equation}

\noindent where {\bf u}$_{\rm QS}$ represents the wave field of the quiet Sun simulation. Thus, the wave field for two non-interacting tubes is

 \begin{equation}
{\bf u}_{ij}^{\rm ss}={\bf u}_{\rm QS}+{\bf u}_{i}^{\rm sc}+{\bf u}_{j}^{\rm sc}
\label{eq:single_scattering}
\end{equation}

\noindent Finally, Hankel analysis is performed on this new set of data. In the following figures, we will compare the results measured for the simulation with two tubes (solid line in top panels) with those obtained from the estimation of non-interacting tubes (dashed line in top panels). The bottom panels will show the difference $\Delta \alpha=\alpha_{\rm ms}-\alpha_{\rm ss}$, where $\alpha_{\rm ms}$ corresponds to the absorption coefficient obtained from the simulation with the two tubes (multiple scattering) and $\alpha_{ss}$ is the absorption measured from ${\bf u}_{ij}^{\rm ss}$ (single scattering).

\subsection{Two tubes: $d=0.5$ Mm, $\chi=0^o$}
\label{sect:d500a0}

\begin{figure}[!ht] 
 \centering
 \includegraphics[width=9cm]{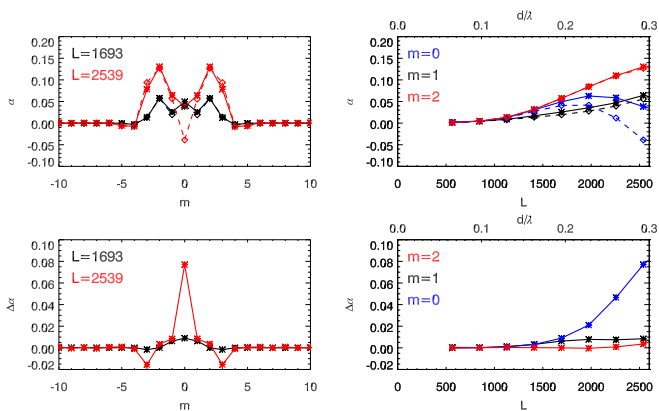}
  \caption{Top panels: Absorption coefficient for the simulation with $d=0.5$ Mm and $\chi=0^o$ (solid line and asterisks), and absorption of two non-interacting tubes with the same configuration (dashed line and diamonds). Bottom panels: Difference between the absorption in the simulation with the two tubes and that of two non-interacting tubes, with positive sign meaning that the latter is smaller. The left panels show the variation with the azimuthal order at $L=1693$ (black line) and at $L=2539$ (red line), while the right panels correspond to the variation with $L$ (bottom axis) or $d/\lambda$ (top axis) at several azimuthal orders.}
  \label{fig:alpha_d500a0}
\end{figure}

The top panels of Figure \ref{fig:alpha_d500a0} show the absorption coefficient $\alpha_m(L)$ for the simulation of two flux tubes with a separation $d=0.5$ Mm located at $y=0$ Mm, that is, the angle $\chi$ is $0^o$. The origin of the coordinate system used for the Hankel analysis is located at the axis of the tube that first encounters the incident wave. The left panel illustrates the distribution of the absorption with the azimuthal order for two example $L$ values. At $L=1693$ it shows peaks at $m=0$ and $m=\pm 2$. However, the decrease of the absorption with $L$ above $L=2000$ (see top right panel) produces a minimum in $m=0$ at $L=2539$. The absorption is significant up to $m=\pm3$, and the two tubes even produce a little emission at $m=\pm4$ and $m=\pm5$. Note that the absorption has perfect symmetry in the sign of the azimuthal order. The variation of the absorption with $L$ for the azimuthal order $m=0$, $m=1$, and $m=2$ can be seen in the top left panel. The latter two increase with $L$, while the absorption for $m=0$ reaches a maximum peak at about $L=2000$ and then decreases.    

In this paper we are considering thin flux tubes. In general, both the distance between tubes and the wavelengths of interest will be much higher than the radius of the tubes. This way, one relevant parameter to characterize the distances is the ratio $d/\lambda$ rather than the absolute distances $d$, where $\lambda$ is the wavelength. In order to help the reader to better appreciate the separations, in those figures where the dependence with $L$ is plotted we have added another axis at the top with the parameter $d/\lambda$.

\begin{figure}[!ht] 
 \centering
 \includegraphics[width=9cm]{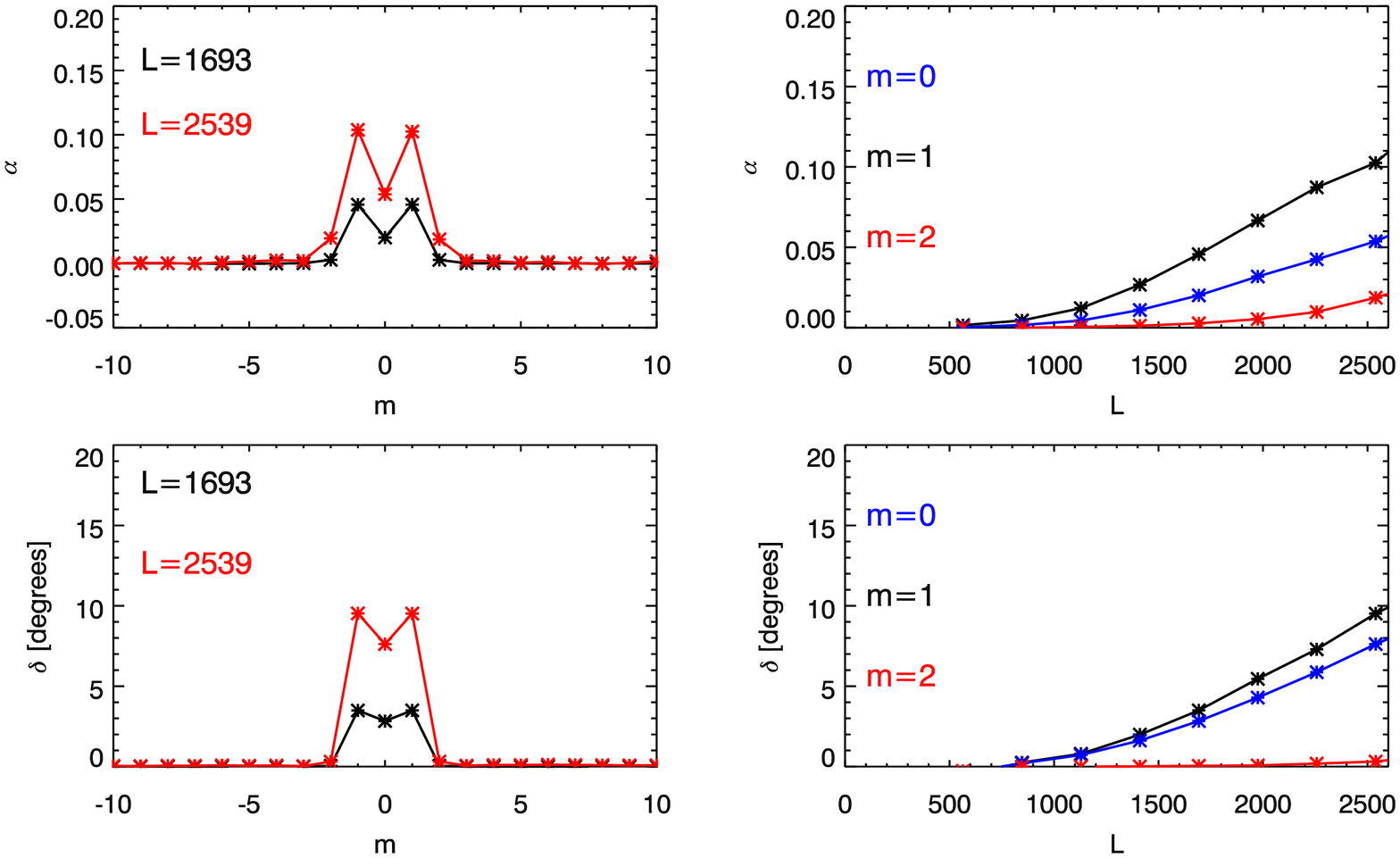}
  \caption{Top panels: Absorption coefficient produced by one tube as a function of azimuthal order (left panel) and $L$ (right panel). Bottom panels: Phase shift produced by one tube as a function of azimuthal order (left panel) and $L$ (right panel).}
  \label{fig:1tube}
\end{figure}

The absorption produced by an isolated tube with the same properties is shown in the top panels of Figure \ref{fig:1tube} for comparison. In this case, the peaks appears at $m=\pm1$, and there is a very little absorption in $m=\pm2$. The total absorption, summing over all azimuthal orders, is clearly higher in the simulation with both tubes, and the distribution in azimuthal orders is significantly different.

As seen in the bottom right panel from Figure \ref{fig:alpha_d500a0}, for harmonic degrees below $L=1411$ multiple scattering effects are negligible. Above this value, the absorption coefficient at $m=0$ produced by the two interacting tubes is much higher than that for the two non-interacting tubes, and the difference increases with $L$. In fact, if multiple scattering effects are neglected, at this azimuthal order the pair of tubes would produce emission, since the absorption coefficient is negative, as can be seen in the top left panel. The emission in the axisymmetric mode is a consequence of the origin of the coordinate system, rather than a real emission. This comparison reveals the relevance of multiple scattering, since  $\Delta \alpha$ is even higher than $\alpha_{ms}$. For the azimuthal orders $m=\pm 1$ and $m=\pm 2$ the absorption coefficient is also increased, but by a much smaller amount than in the case of the axisymmetric mode. Note that $\Delta \alpha (m=1)$ is exactly equal to $\Delta \alpha (m=-1)$. In fact,  $\Delta \alpha$ is perfectly symmetric around $m=0$, as seen in the left panel from Figure \ref{fig:alpha_d500a0}. It shows that multiple scattering increases the absorption for $|m|\le 2$ and a little bit in $m=\pm 5$, while it produces a reduced absorption in $m=\pm 3$.    

For an understanding of the way that the multiple scattering affects the absorption coefficient, some care must be taken regarding the importance of the cylindrical coordinates used for the Hankel analysis in the power distribution of the different azimuthal orders and, thus, the absorption. In an isolated tube case, a cylindrical coordinate system can be assigned at the axis of the tube. Since the tube is axisymmetric, no scattering can be produced from an azimuthal order $m$ to a different order $m'$. When several tubes are present, the symmetry is broken and mode mixing between different azimuthal orders can take place. For example, the power of a purely axisymmetric wave scattered by one tube would be seen as a $m=0$ mode for a coordinate system located at that tube. However, from the point of view of a second tube shifted from the position of the first tube, that scattered wave would consist of a mixture of $m=0$ and higher azimuthal orders, and it would scatter waves in all these orders. Finally, in the coordinate system of the first tube, these waves would be seen as mixture of several azimuthal orders, despite starting with just an $m=0$ mode.

\begin{figure*}[!ht] 
 \centering
 \includegraphics[width=18cm]{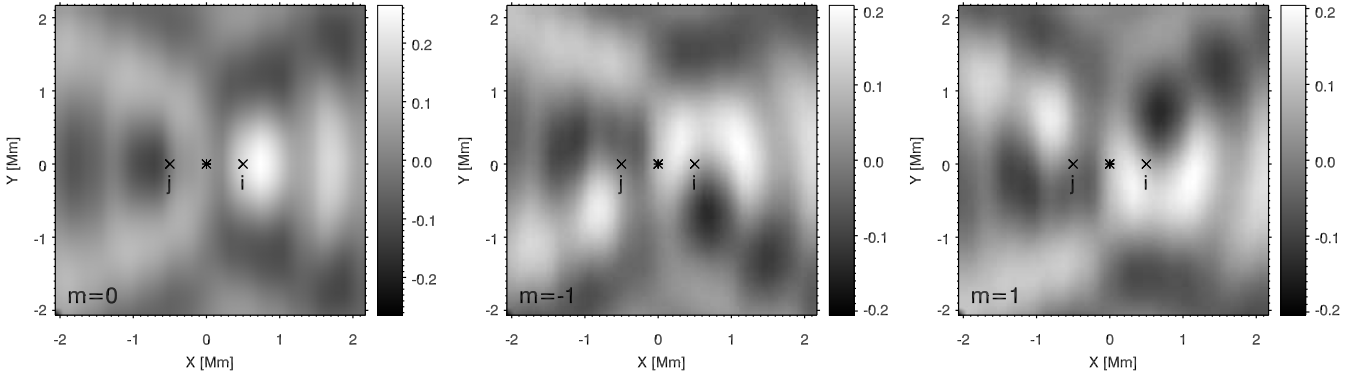}
  \caption{Absorption maps $\alpha_{m}(x,y,L)$ in the azimuthal orders $m=0$ (left panel), $m=-1$ (middle panel), and $m=1$ (right panel) at $L=2539$ for a flux tube located at $x=0$ Mm and $y=0$ Mm (asterisk). The locations at $y=0$ and $x=0.5$ Mm or $x=-0.5$ are indicated by $i$ and $j$, respectively.}
  \label{fig:absorption_maps}
\end{figure*}

Figure \ref{fig:absorption_maps} shows the absorption maps $\alpha_m(x,y,L)$ for the simulation with one tube located at $x=y=0$ in the azimuthal orders $m=0$ and $m=\pm 1$ at $L=2539$. These absorption maps are obtained by repeating the Hankel analysis at all points around the tube. The value at each location is the absorption coefficient in $L=2539$ and the corresponding $m$ as seen by a cylindrical coordinate system located at that position. Note that at some locations the Hankel analysis can see emission from a single flux tube at a certain azimuthal order as a result of a shifted coordinate system. According to Figure \ref{fig:1tube}, at $L=2539$ one isolated tube produces an absorption slightly above 0.05 in $m=0$. This value is obtained by performing the Hankel analysis at the axis of the tube, that is, at the position of the asterisk in Figure \ref{fig:absorption_maps}. However, from the point of view of a coordinate system shifted 0.5 Mm to the right (position $i$ in the figure), as the companion tube in the simulation discussed in this section, the absorption in $m=0$ is significantly higher. It means that the second tube sees a lower outgoing power $|B_{m=0} (L=2539)|^2$ in comparison to the ingoing power $|A_{m=0} (L=2539)|^2$ due to the absorption produced by the first tube. The second flux tube will produce an extra absorption in $m=0$ in turn and will reduce even more the outgoing power $|B_{m=0} (L=2539)|^2$. If now we consider that the asterisk represents the second tube, its absorption will then be seen by the first tube (the position that we have used as center of the Hankel-Fourier decomposition for the analysis of the simulation with two tubes) as absorption at position $j$. All in all, the absorption in $m=0$ produced by two tubes aligned with the direction of propagation of the $f-$mode with a separation of 0.5 Mm will be higher than that for two non-interacting tubes, as seen in Figure \ref{fig:alpha_d500a0}. The multiple scattering effects are produced by an interplay between the scattered wave fields as seen by the companion tubes. 

The absorption maps of higher azimuthal orders show more complex patterns. As an example, the middle and right panels from Figure \ref{fig:absorption_maps} show absorption maps for $m=-1$ and $m=1$, respectively. As can be seen from a comparison of both maps $\alpha_{m=1}(x,y,L)=\alpha_{m=-1}(x,-y,L)$, which means that at $y=0$ Mm the two maps are equal. Thus, multiple scattering effects from a pair of tubes aligned with the direction of propagation of the $f$ mode (located at $i$ and asterisk positions, for example) would be the same in $m=1$ and $m=-1$, as seen in Figure \ref{fig:alpha_d500a0}.

\begin{figure}[!ht] 
 \centering
 \includegraphics[width=9cm]{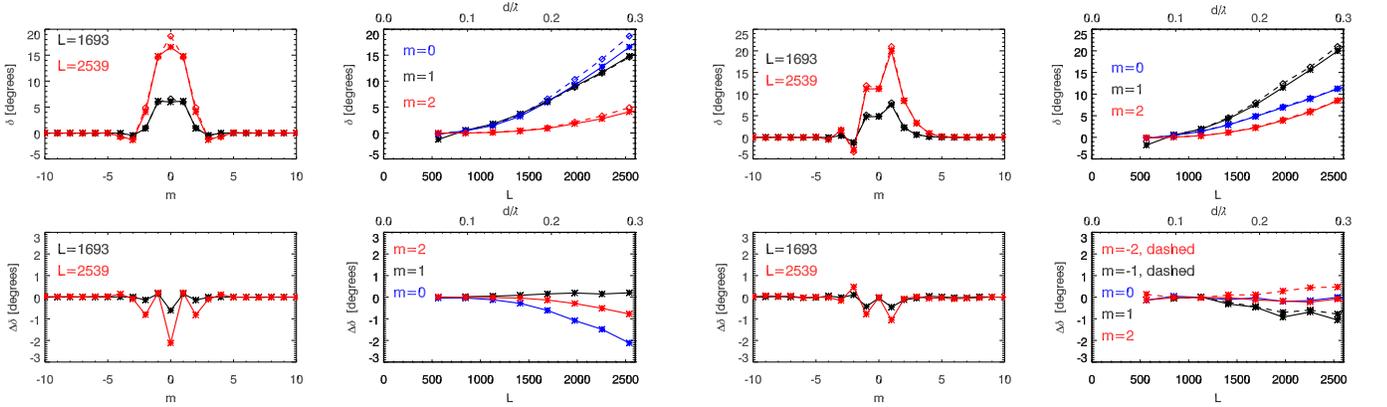}
  \caption{Top panels: Phase shift for the simulation with $d=0.5$ Mm and $\chi=0^o$ (solid line and asterisks), and the phase shift of two non-interacting tubes with the same configuration (dashed line and diamonds). Bottom panels: Difference between the phase shift in the simulation with two tubes and that of two non-interacting tubes. The left panels show the variation with azimuthal order at $L=1693$ (black line) and at $L=2539$ (red line), while the right panels correspond to the variation with $L$ at several azimuthal orders.}
  \label{fig:phase_d500a0}
\end{figure}

Figure \ref{fig:phase_d500a0} shows the phase shift produced by the two tubes separated by 0.5 Mm with $\chi=0^o$, and a comparison with the phase shift that would be generated by a pair of non-interacting tubes located at the same positions. The configuration is the same as in Figure \ref{fig:alpha_d500a0}. As in the case of one tube (bottom panels from Figure \ref{fig:1tube}), the phase shift increases with $L$. However, the highest phase shift is produced in $m=0$, which differs from the case with just one tube where the phase shift peaks at $m=\pm 1$. With two tubes with this configuration, the phase shift at $m=\pm 1$ is slightly lower than that at $m=0$. The effects of multiple scattering tend to reduce the phase shift, especially in $m=0$, where $\Delta \delta$ decreases with $L$ until reaching $\Delta \delta\approx -2.1^o$ at $L=2539$. The distribution of the phase shift with the azimuthal order is also symmetric in $m$.

\subsection{Two tubes: $d=0.5$ Mm, $\chi=90^o$}
\label{sect:d500a90}

\begin{figure}[!ht] 
 \centering
 \includegraphics[width=9cm]{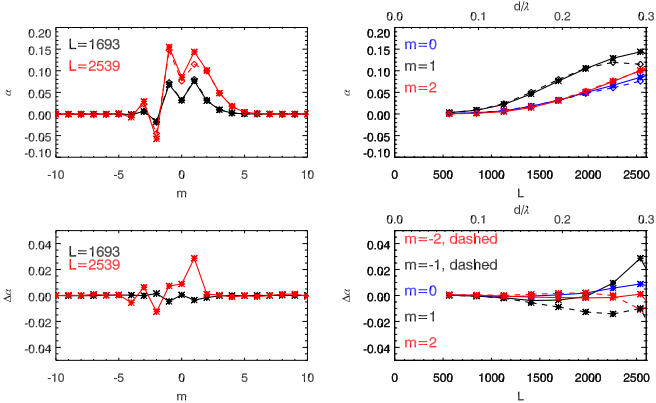}
  \caption{Absorption coefficient for two tubes with $d=0.5$ Mm and $\chi=90^o$, following Figure \ref{fig:alpha_d500a0}. The symmetry in the sign of the azimuthal order is broken due to the presence of the companion tube.}
  \label{fig:alpha_d500a90}
\end{figure}

\begin{figure}[!ht] 
 \centering
 \includegraphics[width=9cm]{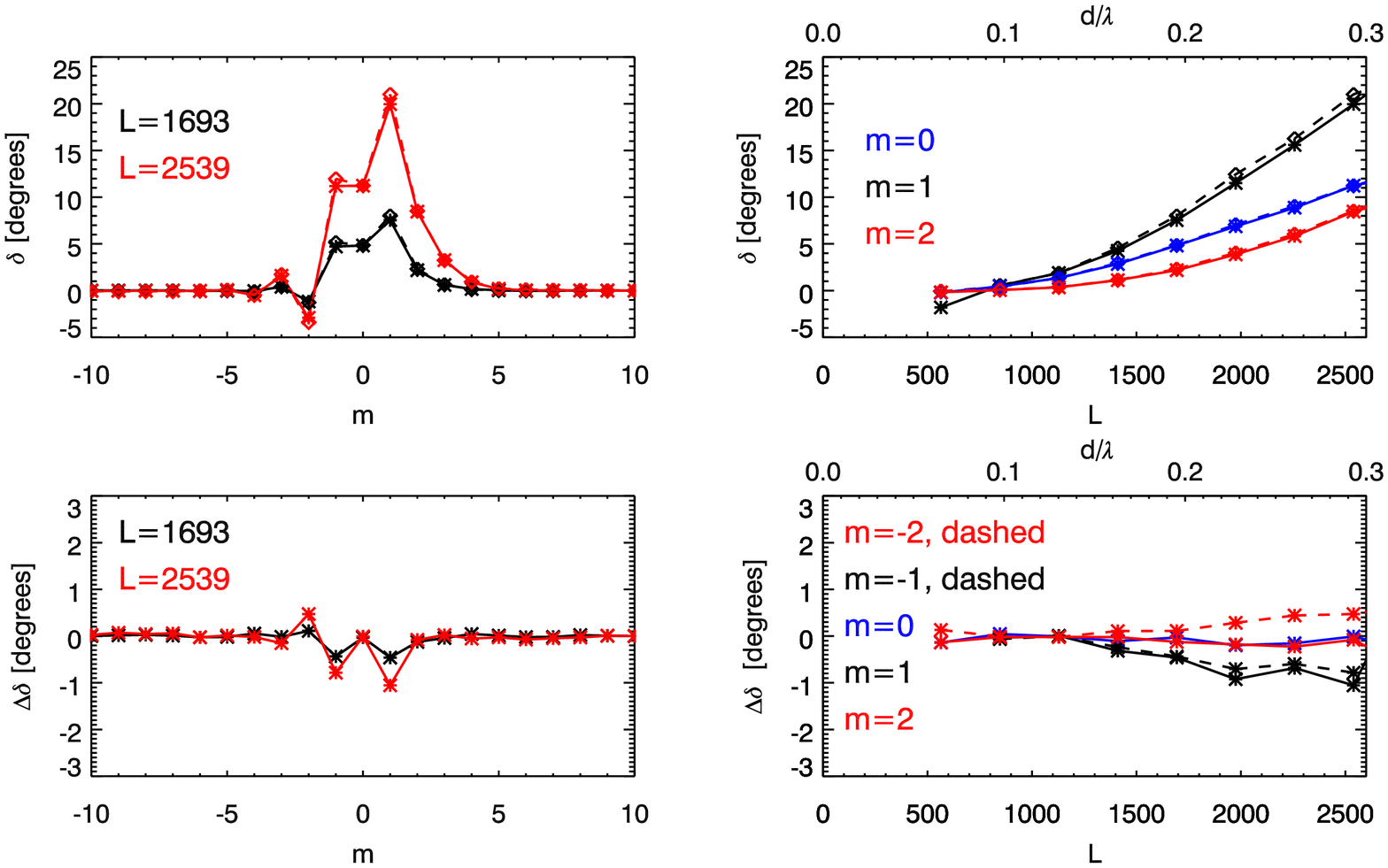}
  \caption{Phase shift for two tubes with $d=0.5$ Mm and $\chi=90^o$, following Figure \ref{fig:phase_d500a0}. The effects of multiple scattering are significantly smaller than in the case with the tubes aligned with the direction of propagation of the incident $f-$mode ($\chi=0^o$, Figure \ref{fig:phase_d500a0})}
  \label{fig:phase_d500a90}
\end{figure}

Figures \ref{fig:alpha_d500a90} and \ref{fig:phase_d500a90} show the absorption and phase shift, respectively, produced by two tubes separated by 0.5 Mm, as in the previous subsection, but in this case $\chi=90^o$. Since the alignment of the two tubes is perpendicular to the direction of propagation of the $f-$mode, the wavefront reaches both tubes at the same time. We have chosen the axis of the tube at $y=-0.25$ Mm as the center of the Hankel analysis. 

The presence of the companion tube breaks the symmetry in the sign of the absorption and phase shift in their distribution with the azimuthal order found for the simulation with one tube (Figure \ref{fig:1tube}), and also for the simulation with two flux tubes aligned with the direction of propagation of the incident wave (Figures \ref{fig:alpha_d500a0} and \ref{fig:phase_d500a0}). The absorption peaks at $m=1$ and $m=-1$, like in the case of just one tube, although in this simulation the latter is slightly higher. The tubes influence the azimuthal orders up to $m=\pm 4$, even producing a significant emission at $m=-2$, and their effects become larger at higher degree $L$. The effects of multiple scattering in the absorption can be seen in the bottom panel of Figure \ref{fig:alpha_d500a90}. In general, multiple scattering effects tend to increase the absorption at high $L$, especially at $m=1$. For the rest of the azimuthal orders, $\Delta \alpha$ shows low positive values, except for $m=-2$ and $m=-4$, which are slightly negative at $L=2539$. Note that the distribution of the power with azimuthal order strongly depends on the position of the coordinate system used for the Hankel-Fourier decomposition. In this way, if the center of the Hankel analysis were located at the axis of the tube at $y=0.25$ Mm, instead of $y=-0.25$ Mm, negative azimuthal orders would show the absorption of the equivalent positive azimuthal order in Figure \ref{fig:alpha_d500a90} and vice versa. On the other hand, if we place the center of the Hankel analysis at the middle position between the two tubes, the results will show symmetry in azimuthal order. 
   
With regards to the phase shift (Figure \ref{fig:phase_d500a90}), multiple scattering produces small differences, only significant at $m=1$ and $m=-1$. Although the difference between the interacting and non-interacting tubes are large at high $L$, it is less than -1.3$^o$ for all the azimuthal orders considered. The maximum phase shift is obtained for $m=1$, which corresponds approximately to twice the phase shift produced by an isolated tube at this azimuthal order (Figure \ref{fig:1tube}).

\subsection{Two tubes: $d=0.5$ Mm, $\chi=45^o$}
\label{sect:d500a45}

\begin{figure}[!ht] 
 \centering
 \includegraphics[width=9cm]{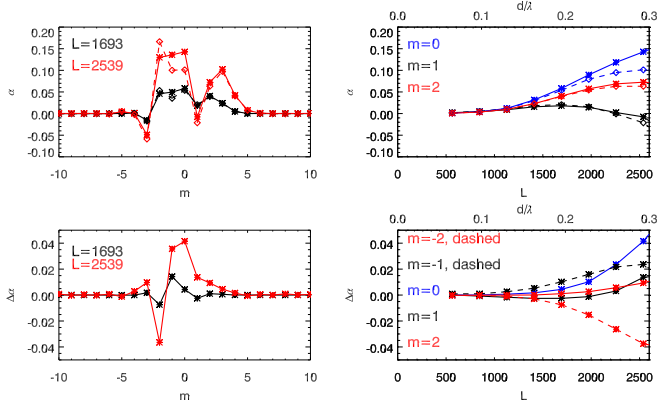}
  \caption{Absorption coefficient for two tubes with $d=0.5$ Mm and $\chi=45^o$, following Figure \ref{fig:alpha_d500a0}. The pair of tubes can show acoustic emission for some azimuthal orders.}
  \label{fig:alpha_d500a45}
\end{figure}

In this case we consider an intermediate situation between the two previous simulations, $\chi=45^o$. As illustrated in Figure \ref{fig:alpha_d500a45}, the highest absorption is found for $m=-2$, $m=-1$, and $m=0$, and it presents remarkable minimums at $m=-3$ and $m=1$, where there is emission. Multiple scattering produces a strong reduction of the absorption in $m=-2$, but in the other azimuthal orders where it is significant the absorption is increased, especially at $m=-1$ and $m=0$. Higher $L$ values are more affected by multiple scattering, as shown by the difference $\Delta\alpha$ in the bottom right panel.

\begin{figure}[!ht] 
 \centering
 \includegraphics[width=9cm]{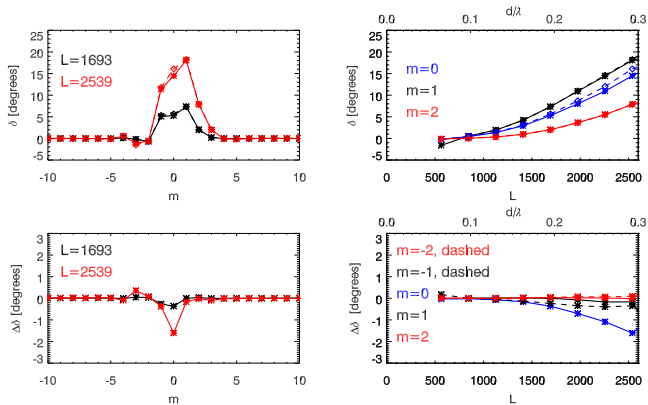}
  \caption{Phase shift for two tubes with $d=0.5$ Mm and $\chi=45^o$, following Figure \ref{fig:phase_d500a0}. This case represents an intermediate situation between the results shown in Figures \ref{fig:phase_d500a0} and \ref{fig:phase_d500a90}.}
  \label{fig:phase_d500a45}
\end{figure}

As in the case of the simulation presented in Section \ref{sect:d500a0}, the phase shift (Figure \ref{fig:phase_d500a45}) is concentrated in $m=0$ and $m=\pm 1$, although in this case the distribution is shifted toward positive azimuthal orders, showing a peak at $m=1$. It increases almost linearly with $L$, and only $m=0$ shows significant differences relative to the non-interacting case. In this azimuthal order at $L=2539$ the multiple scattering produces a reduction of around $1.6^o$ of the phase shift. A comparison of the phase shift in $m=0$ for the three simulations with $d=0.5$ Mm and different angle $\chi$ reveals that the largest reduction of the phase shift is produced for $\chi=0^o$. As $\chi$ increases the effects of the multiple scattering in the axisymmetric phase shift are reduced, and they are negligible for $\chi=90^o$.

\subsection{Two tubes: $d=1$ Mm, $\chi=0^o$}
\label{sect:d1000a0}

\begin{figure} 
 \centering
 \includegraphics[width=9cm]{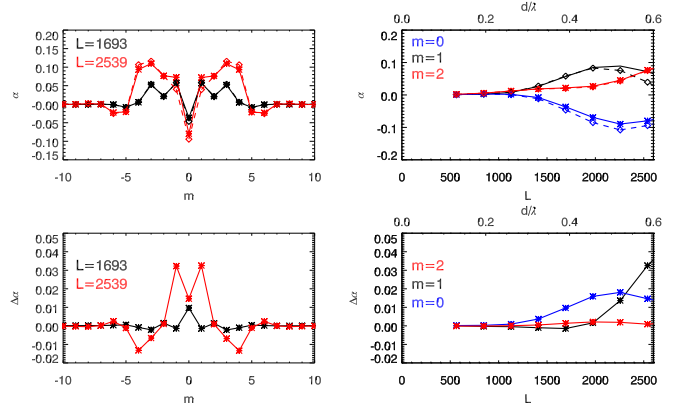}
  \caption{Absorption coefficient for two tubes with $d=1$ Mm and $\chi=0^o$, following Figure \ref{fig:alpha_d500a0}. Since the tubes are aligned with the direction of propagation of the incident wave, the absorption coefficient is symmetric in azimuthal order.}
  \label{fig:alpha_d1000a0}
\end{figure}

\begin{figure}
 \centering
 \includegraphics[width=9cm]{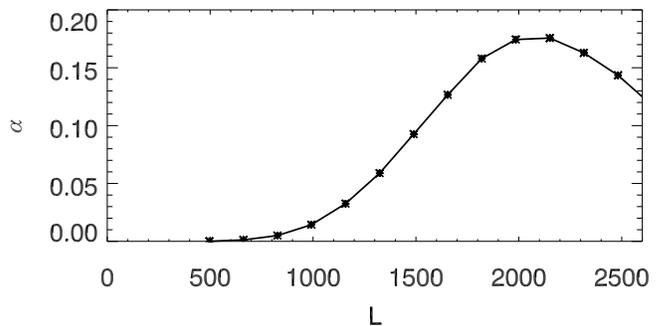}
  \caption{Variation of the absorption coefficient with $L$ at $m=0$ produced by one isolated tube. The coordinate system used for the Hankel analysis is centered at $x=1$ Mm and $y=0$ (the flux tube is located at $x=y=0$).}
  \label{fig:alpha_L_x1000}
\end{figure}

In this subsection and the following we aim to evaluate the coupling between two tubes at different separations, and compare with the case previously presented in Section \ref{sect:d500a0}. In this simulation we have located the companion tube at 1 Mm from the first tube, and the two tubes are aligned with the direction of propagation of the $f$ mode. Figure \ref{fig:alpha_d1000a0} shows the absorption coefficient as seen by a coordinate system placed at the axis of the tube that first encounters the incident wave, as usual. The absorption shows a prominent minimum at $m=0$, similar to the case with the tubes separated by 0.5 Mm (Figure \ref{fig:alpha_d500a0}). However, in this case it corresponds to negative absorption, which means that due to the second tube, the tube that first encounters the $f-$mode sees emission in the axisymmetric mode. At $L=2539$ the absorption at azimuthal orders from 1 through 4 (and the negative counterparts) is between 0.07 and 0.11, while at higher orders it almost vanishes. As expected, the distribution in azimuthal order is symmetric. Multiple scattering tends to increase the absorption for azimuthal orders with $|m|<1$, while it reduces the absorption at $|m|=3$ and $|m|=4$. As we stated before, the effects of multiple scattering on the absorption are produced by the variations in the wave field that each tube sees due to the presence of the companion tube. As an example, we are going to focus on $m=0$. The position of the monopole absorption peak produced by an isolated tube (Figure \ref{fig:absorption_maps}, left panel) changes with the wavelength, and it is located closer to the tube at high $L$. Figure \ref{fig:alpha_L_x1000} shows the variation of the absorption coefficient with $L$ produced by one isolated tube as seen from a distance of 1 Mm and aligned with the direction of propagation of the incident $f-$mode. The highest absorption is obtained around $L=2000$. An observer located at that position would see the highest difference between the incoming power and the outgoing power at that $L$, with the former being higher. Comparing Figure \ref{fig:alpha_L_x1000} with the bottom right panel of Figure \ref{fig:alpha_d1000a0}, we found that the strongest multiple scattering effects in $m=0$ are also obtained at that $L$. 

\begin{figure}[!ht] 
 \centering
 \includegraphics[width=9cm]{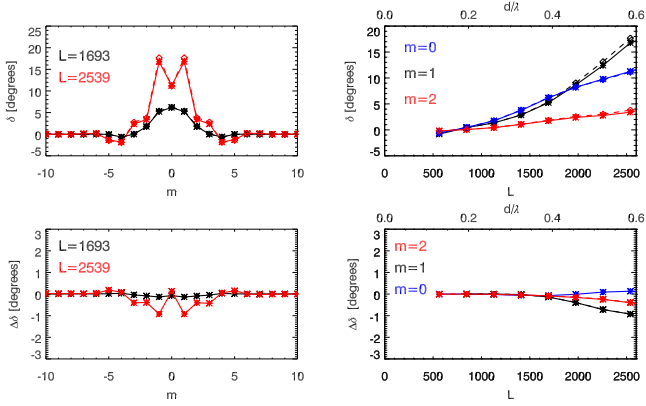}
  \caption{Phase shift for the simulation with $d=1$ Mm and $\chi=0^o$, following Figure \ref{fig:phase_d500a0}. Multiple scattering effects have a small contribution to the phase shift.}
  \label{fig:phase_d1000a0}
\end{figure}

The phase shift (Figure \ref{fig:phase_d1000a0}) shows two peaks at $m\pm 1$ with a minimum at $m=0$ for $L=2539$. This situation is significantly different from the case of two tubes separated by 0.5 Mm (Figure \ref{fig:phase_d500a0}), where we obtained just one peak at $m=0$. The larger separation produces a detachment of the two peaks, and the phase shift is more similar to that produced by an isolated tube (Figure \ref{fig:1tube}). However, due to the wave field of the companion tube, a small phase shift is measured at higher azimuthal orders, up to $m=\pm 5$. Multiple scattering produces a reduction of the phase shift for most azimuthal orders and degrees, and its effects are more significant at higher $Ls$.

\subsection{Two tubes: $d=4$ Mm, $\chi=0^o$}
\label{sect:d4000a0}

\begin{figure}
 \centering
 \includegraphics[width=9cm]{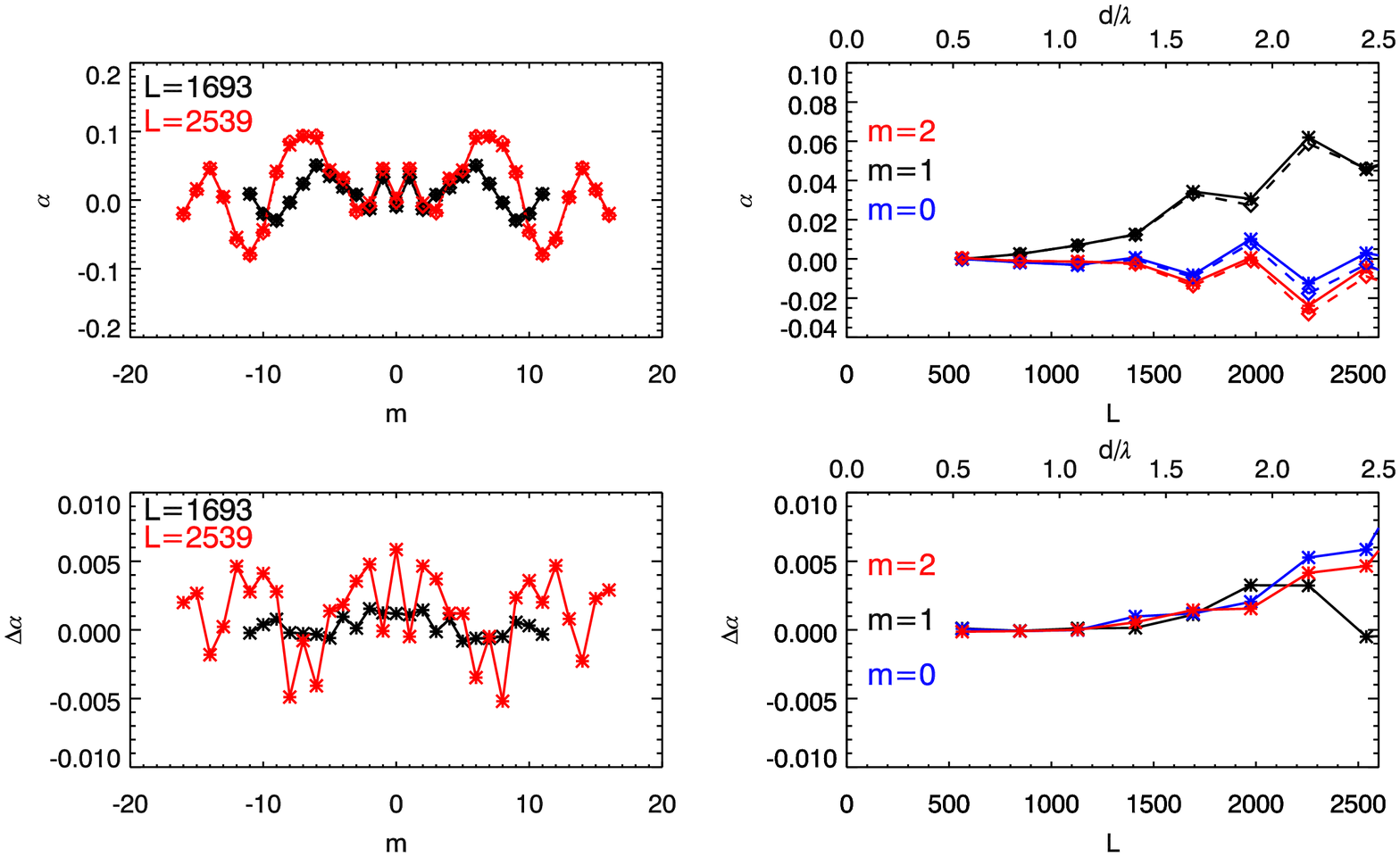}
  \caption{Absorption coefficient for two tubes with $d=4$ Mm and $\chi=0^o$, following Figure \ref{fig:alpha_d500a0}. Note the different scales in the bottom panels. Multiple scattering effects are negligible.}
  \label{fig:alpha_d4000a0}
\end{figure}

\begin{figure}[!ht] 
 \centering
 \includegraphics[width=9cm]{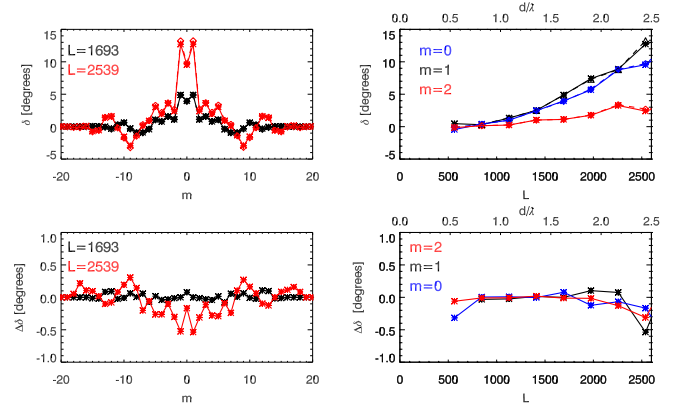}
  \caption{Phase shift for the simulation with $d=4$ Mm and $\chi=0^o$, following Figure \ref{fig:phase_d500a0}. Note the different scales. Multiple scattering effects are negligible.}
  \label{fig:phase_d4000a0}
\end{figure}

Figures \ref{fig:alpha_d4000a0} and \ref{fig:phase_d4000a0} show the results for another case with two tubes aligned with the propagation of the $f-$mode, but located at a higher separation, in this case at 4 Mm. A comparison with the absorption produced by pairs of tubes with smaller separations, as those shown in previous sections, reveals significant differences. In this case high absorption (and emission) is found for high azimuthal orders, up to $m=\pm 17$, and the absorption shows a complex distribution with azimuthal order. Moreover, for all the azimuthal orders plotted in the top right panel of Figure \ref{fig:alpha_d4000a0}, the absorption oscillates with the degree $L$. However, this behavior is not produced by the interaction of both tubes. A similar pattern is found for the absorption produced by a pair of non-interacting tubes at those positions (dashed lines from top panels of Figure \ref{fig:alpha_d4000a0}). The distribution in $m$ and $L$ is mainly caused by how the first tube sees the wave field produced by the second tube. The effects of multiple scattering are shown by the bottom panels of Figure \ref{fig:alpha_d4000a0}. The variations of the absorption produced by a second tube located at this distance are negligible. 

The phase shift (Figure \ref{fig:phase_d4000a0}) shows two peaks at $m\pm 1$ with a minimum at $m=0$ for $L=2539$, equivalent to an isolated tube and two tubes with a separation of 1 Mm. Similar to the latter case, the second tube generates a small phase shift at high azimuthal orders, but in this case it is measurable at even higher azimuthal orders, up to $m=\pm 16$. The second tube also produces some changes in the variation of $\delta$ with $L$. While for one tube it increases almost linearly for all azimuthal orders, in this case there are small variations with respect to a linear increase. Multiple scattering effects in the phase shift for two tubes separated by 4 Mm is almost negligible (below 0.5$^o$ for all azimuthal orders).

\subsection{Three tubes: $d=0.5$ Mm, $\chi=0^o$}
\label{sect:3tubes}

In the previous sections, we have characterized the absorption and phase shift for a bundle of two tubes in several configurations, varying the distance and the angle between the tubes, and comparing the results with those expected for non-interacting tubes. As a next step we have performed a simulation with three flux tubes, all of them aligned with the direction of propagation of the $f$ mode and with a separation of 0.5 Mm between them. There are two objectives for this simulation. On the one hand, we aim to evaluate how the measures of the absorption and phase shift are modified by the addition of more flux tubes to the bundle. On the other hand, in the previous two tubes simulations we have shown that single scattering cannot explain the measurements, since multiple scattering effects are significant. However, can they be approximated by two interactions? In order to evaluate the absorption and phase shift for three aligned tubes including two interactions scattering, we have used a set of six simulations, most of them already used in the previous sections, together with the quiet Sun simulation. This set includes three simulations with only one tube being present at $x=-0.5$, $x=0$, or $x=0.5$ Mm (we will refer to the wave fields from this simulations as {\bf u}$_1$, {\bf u}$_2$, and {\bf u}$_3$, respectively), and other three simulations with two flux tubes, covering all the possible pairs of tubes at those spatial position (in the case of the simulation with flux tubes located at $x=-0.5$ and $x=0$ the wave field will be indicated as {\bf u}$_{12}$, and following the same criteria the wave field of the other two simulations will be {\bf u}$_{13}$ and {\bf u}$_{23}$). The wave field of the scattered wave of these simulations with two tubes can be written as

 \begin{equation}
{\bf u}_{ij}^{\rm sc}={\bf u}_{i}^{\rm sc}+{\bf u}_{j}^{\rm sc}+\Delta {\bf u}_{ij}
\label{eq:twotubes}
\end{equation}

\noindent where $\Delta {\bf u}_{ij}$ represents the variation in the wave field due to multiple scattering effects between tubes $i$ and $j$. The wave field for three tubes including two interactions(superscript $2i$) is obtained as

\begin{eqnarray}
\label{eq:u123}
\lefteqn{{\bf u}_{123}^{\rm 2i}={\bf u}_{QS}+{\bf u}_{12}^{\rm sc}+{\bf u}_{23}^{\rm sc}+{\bf u}_{13}^{\rm sc}-{\bf u}_{1}^{\rm sc}-{\bf u}_{2}^{\rm sc}-{\bf u}_{3}^{\rm sc}} \nonumber\\
&&={\bf u}_{QS}+{\bf u}_{1}^{\rm sc}+{\bf u}_{2}^{\rm sc}+{\bf u}_{3}^{\rm sc}+\Delta {\bf u}_{12}+\Delta {\bf u}_{23}+\Delta {\bf u}_{13}.
\end{eqnarray}

As shown in the previous equation, ${\bf u}_{123}^{\rm 2i}$ includes the individual contribution of each tube to the wave field as well as the multiple scattering produced by each pair of tubes. As we did in previous sections, we have also quantified the absorption and phase shift for three non-interacting tubes using the velocity wave field from the three simulations with an isolated tube.

\begin{figure}
 \centering
 \includegraphics[width=9cm]{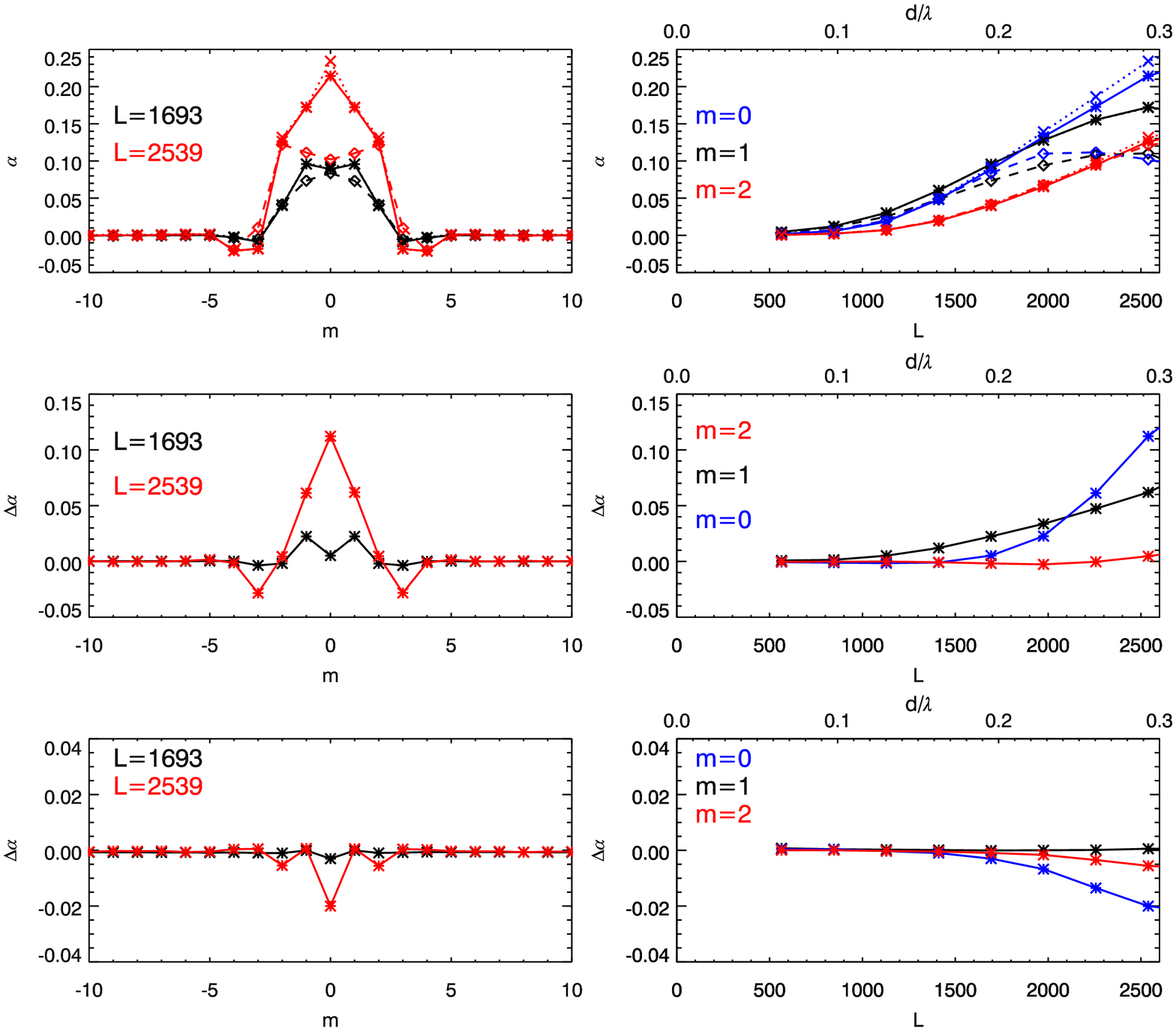}
  \caption{Top panels: Absorption coefficient for three tubes with $d=0.5$ Mm and $\chi=0^o$ (solid line and asterisks), three non-interacting tubes with the same configuration (dashed line and diamonds), and three tubes including interactions of pairs of tubes (dotted line and crosses). Middle panels: Difference between the absorption in the simulation with three tubes and that of three non-interacting tubes. Bottom panels: Difference between the absorption in the simulation with three tubes and 
that of three tubes including interactions of pairs of tubes. The left panels show the variation with the azimuthal order at $L=1693$ (black line) and at $L=2539$ (red line), while the right panels correspond to the variation with $L$ at several azimuthal orders. Note that the scale is different in the second two rows.}
  \label{fig:alpha_3tubes}
\end{figure}

Figure \ref{fig:alpha_3tubes} shows the absorption coefficient for the three tubes. We have located the coordinate center of the Hankel analysis in the middle tube. Azimuthal orders with $|m|<2$ show significant absorption, with a clear peak at $m=0$, while $|m|=4$ and $|m|=5$ present a small amount of emission. The dashed line represents the absorption for the case of non-interacting tubes. At $L=2539$ it is very clear that multiple scattering effects produce a significant increase of the absorption at $m=0$ and $m=\pm 1$. The most interesting result from the analysis of this simulation is the comparison of the absorption coefficient produced by three tubes and that expected when only two interactions are considered (dotted line and crosses). The latter was obtained by performing the Hankel-Fourier analysis on ${\bf u}_{123}^{2i}$. The agreement between the absorption coefficient obtained from ${\bf u}_{123}$ (the simulation the three tubes) and that retrieved from ${\bf u}_{123}^{2i}$ is remarkable. Thanks to the inclusion of two-interaction scattering, the absorption coefficient shows a peak at $m=0$, which was absent when we considered only single scattering. The bottom panels show the discrepancies between both aproximations and the real absorption, revealing a great improvement in the two-interaction scattering case. Only the absorption coefficient for $m=0$ at high $L$ departs significantly from the exact result.

\begin{figure}
 \centering
 \includegraphics[width=9cm]{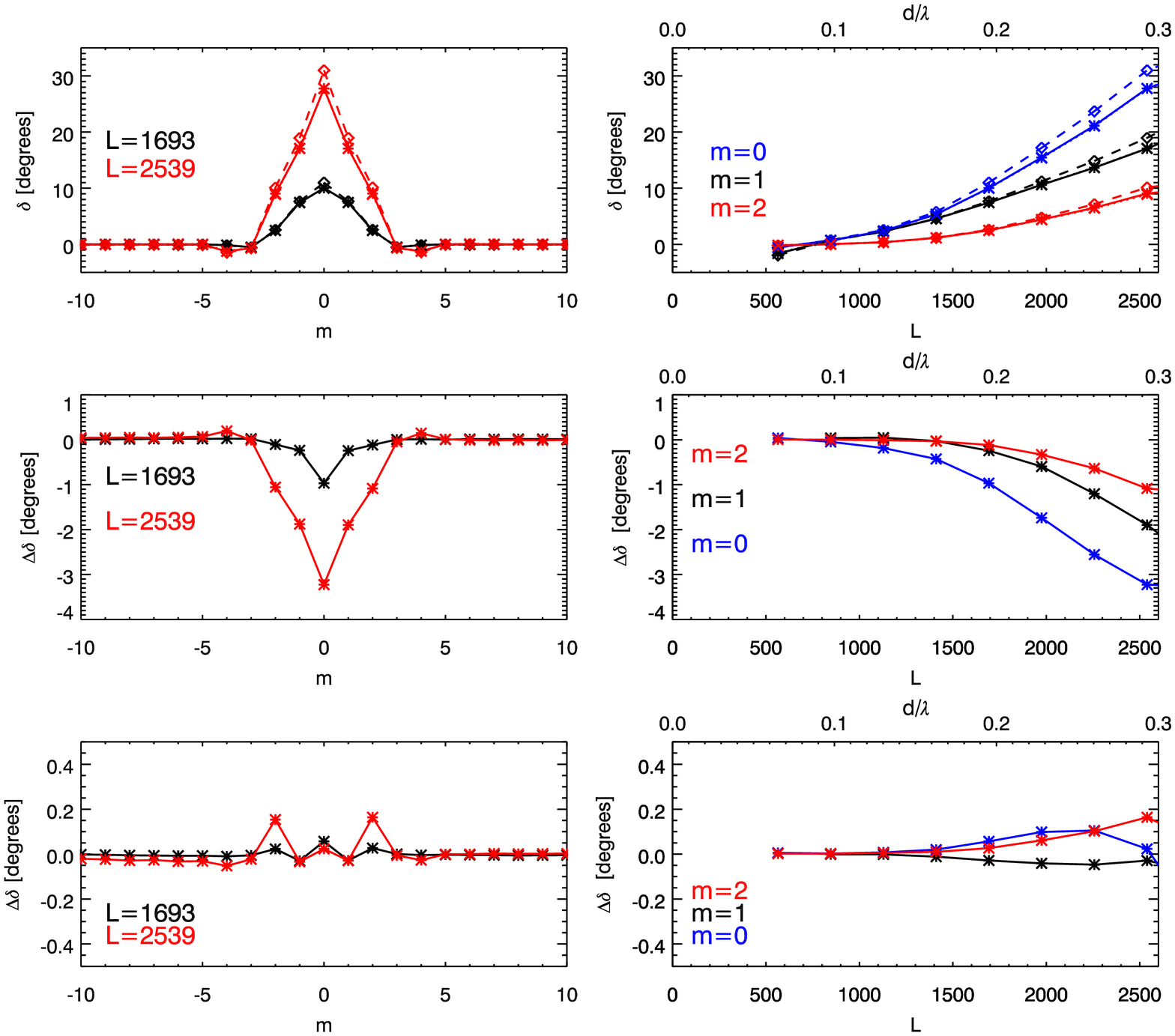}
  \caption{Top panels: Phase shift for three tubes with $d=0.5$ Mm and $\chi=0^o$ (solid line and asterisks), three non-interacting tubes with the same configuration (dashed line and diamonds), and three tubes including interactions of pairs of tubes (dotted line and crosses). Middle panels: Difference between the phase shift in the simulation with three tubes and that of three non-interacting tubes. Bottom panels: Difference between the phase shift in the simulation with three tubes and that of three tubes including interactions of pairs of tubes. The left panels show the variation with the azimuthal order at $L=1693$ (black line) and at $L=2539$ (red line), while the right panels correspond to the variation with $L$ at several azimuthal orders. Note that the scale is different in the second two rows.}
  \label{fig:phase_3tubes}
\end{figure}

With regards to the phase shift produced by the bundle of three flux tubes (Figure \ref{fig:phase_3tubes}), the result is similar to that obtained for two flux tubes also aligned with the propagation of the incident wave (Figure \ref{fig:phase_d500a0}), but in this case the peak at $m=0$ is more prominent. The results for the simulation with the three tubes and the estimations of two-interaction scattering are almost indistinguishable, while including single scattering produces an increased phase shift. These differences can be compared by looking at the bottom panels of the figure. As shown previously for the absorption coefficient, the phase shift of three interacting flux tubes can be better estimated by taking into account the interactions of pairs of tubes. The agreement is striking for almost all $Ls$ and azimuthal orders, only at high $L$ do some discrepancies arise.

\subsection{Ensembles of flux tubes}
\label{sect:14random}

After evaluating the multiple scattering effects produced by two and three interacting tubes, we now study a more general case. We have performed five simulations with different ensembles of flux tubes, all of them with the same properties described in Section \ref{sect:procedures}. The distributions of the tubes were chosen in order to study the impact of two parameters on the results: the average distance between tubes and the number of constituent elements. The former aim is evaluated by means of three simulations with fourteen tubes with different average distances between tubes, defined as $d_{\rm av}=\sum_{i,j}d_{ij}/n^2$, where $d_{ij}$ is the distance between tubes $i$ and $j$, and $n$ is the number of tubes. The positions of the tubes in the most tightly packed ensemble were obtained by placing the fourteen tubes inside a radial distance of 1 Mm centered at $x=y=0$. A minimum distance of 0.5 Mm between the centers of two neighboring tubes was imposed. For this collection $d_{\rm av}=1.07$ Mm. The other two ensembles were constructed by increasing the radial distance of the tubes from the center $x=y=0$. Their average distance is $2.14$ and $3.20$ Mm. The center of the annular region used for the Hankel analysis is located at $x=y=0$.

\begin{figure} 
 \centering
 \includegraphics[width=9cm]{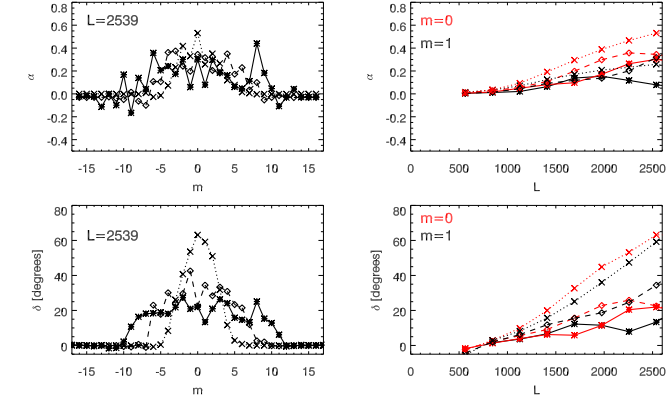}
  \caption{Absorption coefficient and phase shift for ensembles of 14 flux tubes with different average distance between tubes: 1.07 Mm (dotted line and crosses), 2.14 Mm (dashed line and diamonds), and 3.14 Mm (solid line and asterisks).}
  \label{fig:alpha_phase_dij}
\end{figure}

\begin{figure}
 \centering
 \includegraphics[width=9cm]{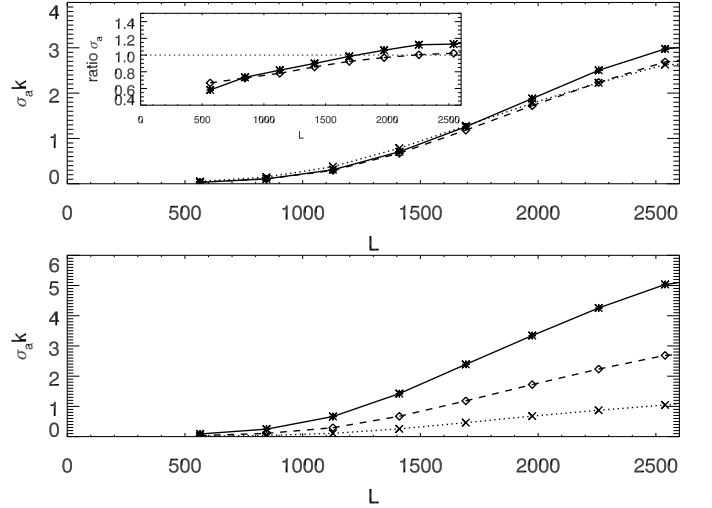}
  \caption{Absorption cross sections (scaled by $k$) for ensembles of tubes. Top panel: ensembles with different average distances between tubes, 1.07 Mm (dotted line and crosses), 2.14 Mm (dashed line and diamonds), and 3.14 Mm (solid line and asterisks). Top panel inset: ratios $\sigma_{\rm a}^{d_{\rm av}=3.20}(L)/\sigma_{\rm a}^{d_{\rm av}=1.07}(L)$ (solid line and asterisks) and $\sigma_{\rm a}^{d_{\rm av}=2.14}(L)/\sigma_{\rm a}^{d_{\rm av}=1.07}(L)$ (dashed line and diamonds). Bottom panel: ensembles with different number of tubes, 7 (dotted line and crosses), 14 (dashed line and diamonds), and 28 (solid line and asterisks).}
  \label{fig:sigma_ensemble}
\end{figure}

Figure \ref{fig:alpha_phase_dij} shows the results for the absorption coefficient and phase shift for these three simulations. The distribution of the absorption coefficient as a function of the azimuthal order is broader for the looser cases ($d_{\rm av}=2.14$ Mm and $d_{\rm av}=3.20$ Mm). This dependence is even clearer in the plot of the phase shift and provides information about the horizontal extent of the absorbing region \citep{Braun+etal1988}, which is obviously larger when the tubes are further apart.  We are interested in evaluating how the distance between the tubes effects the scattering produced by the ensemble. One way to characterize the interaction of the acoustic waves with the group of tubes is the absorption cross section \citep{Keppens+etal1994}, defined as

\begin{equation}
\label{eq:fmode}
\sigma_{\rm a}(L)=\frac{1}{k}\sum_m \alpha_m(L),
\end{equation}

\noindent where $k$ is the horizontal wavenumber. The top panel of Figure \ref{fig:sigma_ensemble} shows $\sigma_{\rm a}(L)$ for the three simulations with fourteen tubes. For degree $L$ below 1700, the simulation with the smallest $d_{\rm av}$ shows the highest absorption cross section. These longer wavelength waves are strongly coupled through their near field, and the closer the tubes the more effective multiple scattering is. At higher $L$, however, the ensemble with larger distance between tubes has the highest absorption cross section. The reason for this is not clear, but it may be related to the larger spatial extent in which the absorbing elements are distributed and the absorption at higher azimuthal orders that it produces. For a better representation of these variations, see the inner plot in the top left corner, where the ratios $\sigma_{\rm a}^{d_{\rm av}=3.20}(L)/\sigma_{\rm a}^{d_{\rm av}=1.07}(L)$ and $\sigma_{\rm a}^{d_{\rm av}=2.14}(L)/\sigma_{\rm a}^{d_{\rm av}=1.07}(L)$ are plotted.

\begin{figure}[!ht] 
 \centering
 \includegraphics[width=9cm]{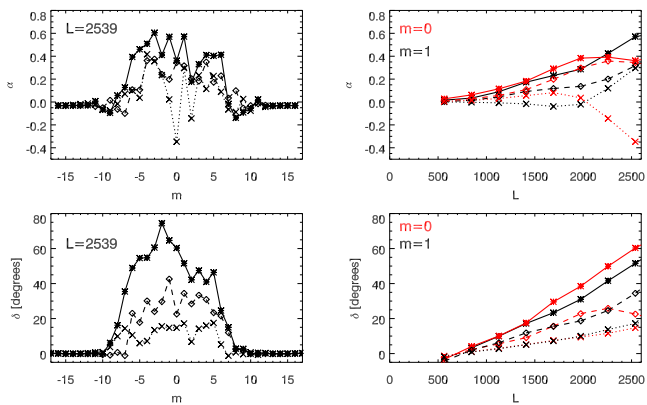}
  \caption{Absorption coefficient and phase shift for ensembles of flux tubes distributed randomly with 7 (dotted line and crosses), 14 (dashed line and diamonds), and 28 (solid line and asterisks) constituent elements. The average distance between tubes is $d_{\rm av}=2.14$ Mm.}
  \label{fig:alpha_phase_ntubes}
\end{figure}

As a next step, we have analyzed the impact of the total magnetic flux in ensembles composed by identical flux tubes, that is, we have compared the results for simulations with different number of tubes. Figure \ref{fig:alpha_phase_ntubes} represents the results for three simulations with seven, fourteen, and twenty-eight flux tubes, all of them with the tubes randomly located but with the same average distance between flux tubes $d_{\rm av}=2.14$ Mm. As expected, adding more tubes produces a significant increase of the absorption and phase shift. There are strong differences in the distribution of these coefficients as a function of azimulthal order. For example, the case with seven tubes shows emission at $m=0$. However, this distribution has a strong dependence on the position of the tubes and how the coordinate system used for the Hankel analysis sees the wavefield produced by them. A distinct configuration would produce a significantly different variation of absorption and phase shift with azimuthal order. 

The bottom panel of Figure \ref{fig:sigma_ensemble} shows the absorption cross section for these three ensembles of tubes. It increases almost linearly with $L$ in the three cases. However, a comparison between them reveals that the absorption cross section is not proportional to the number of tubes. Table \ref{tab:sigma_a} shows the ratio between the absorption cross sections $\sigma_{\rm a}^n(L)$ of the ensembles of tubes for two different $Ls$, where $n$ indicates the number of tubes. The ratio at $L=564$ is always higher than the ratio obtained for $L=2539$, since at longer wavelengths the parameter characterizing the average distance between tubes $d_{\rm av}/\lambda$ is smaller and the tubes are more strongly coupled. In general, the increase in the absorption produced by groups with more tubes is higher than that expected from a proportional increase according to the magnetic flux (\ie, number of tubes), due to the extra interactions produced by the presence of more tubes. The only exception is the comparison between the ensembles with fourteen and twenty-eight tubes at high $L$.





\begin{table}
\begin{center}
\caption[]{\label{tab:sigma_a}
          {Ratio between the absorption cross section of the ensembles of tubes for $L=564$ and $L=2539$}}
\begin{tabular*}{8cm}{@{\extracolsep{\fill}}cccc}

\hline   &    $\sigma_{\rm a}^{14}/\sigma_{\rm a}^7$   &  $\sigma_{\rm a}^{28}/\sigma_{\rm a}^7$   & $\sigma_{\rm a}^{28}/\sigma_{\rm a}^{14}$  \\
 \hline
 L=564   &  2.74      &  6.40   &  2.32 \\
 L=2539  & 2.56       &   4.81     &  1.87  \\

\hline

\end{tabular*}
\end{center}
\end{table}

\begin{figure}[!ht] 
 \centering
 \includegraphics[width=9cm]{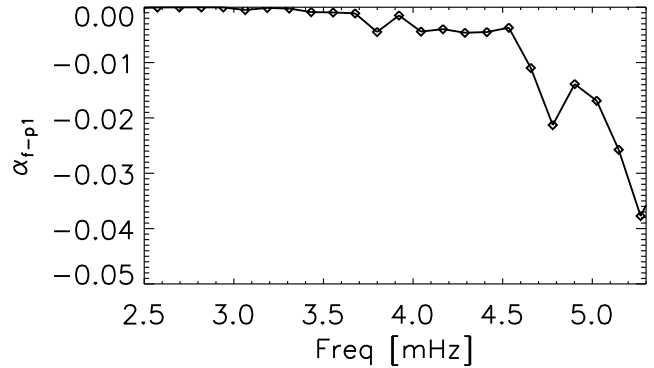}
  \caption{Variation of the mode mixing for all azimuthal orders from the $f-$mode to the $p_1$ mode with frequency for the simulation with fourteen tubes randomly distributed and $d_{\rm av}=1.07$ Mm.}
  \label{fig:mixing_14tubes}
\end{figure}

During the interaction between the flux tubes and the waves, part of the incident $f-$mode power is dispersed into outgoing $p-$modes with the same frequency as the incident wave but with different degree $L$, through a process called mode mixing \citep[\eg,][]{D'Silva1994, Braun1995}. In the simulations with two and three flux tubes, the amount of $p$ mode power generated almost vanishes, but the cases with more tubes scatters measurable power into the $p_1$ ridge. As an example, we have evaluated the mode mixing for all azimuthal orders in the simulation with fourteen tubes and $d_{\rm av}=1.07$ Mm. In this simulation the incident wave is only composed of an $f-$mode, and since the scatterer only produces a significant amount of power in the $p_1$ ridge, we restrict our analysis to mode mixing from the $f$ mode to the $p_1$ mode. In order to quantify the mode mixing, we used the same coefficient $\alpha_{f-p1}(\nu)$ defined in Equation 17 from \citet{Felipe+etal2012a}. It corresponds to the ratio between the outgoing power in the $p_1$ ridge and the incoming power in the $f$ mode with a negative sign, so a negative value indicates emission to $p_1$. Figure \ref{fig:mixing_14tubes} shows the variation of the coefficient $\alpha_{f-p1}$ with frequency. At low frequencies the group of tubes produces little emission in the $p_1$ mode, but mode mixing becomes significant around 5 mHz. For a comparison with flux tube models with different radius see Figure 10 from \citet{Felipe+etal2012a}.

\section{Discussion and conclusions}
\label{sect:conclusions}

We have studied the absorption and phase shift of an $f$ mode produced by groups of flux tubes, focusing on the effects of multiple scattering and its comparison with what one would expect for equivalent cases but neglecting these effects or partially including them. With this aim, we have performed several numerical simulations where we analyzed different situations, from the scattering produced by a pair of tubes, to that of a more general case where a randomly distributed group of tubes (up to 28) are embedded in the atmosphere. In the simulations of two interacting tubes, we have evaluated several configurations in order to quantify how variations in the distance between the tubes and the angle between the line joining their centers and the direction of propagation of the incident $f$ mode modify the properties of the scattered waves. The absorption and phase shift measured for these simulations are compared with those obtained from the sum of the contributions of the same number of tubes located at equivalent positions, but retrieved from simulations with only one tube being present and, thus, representing the absorption and phase shift generated by non-interacting tubes. This method allows us to isolate the multiple scattering effects produced by the interaction of the tubes, which can be retrieved as the difference between the measures for the simulation with several tubes and those for an equivalent case obtained by summing isolated tubes.

\begin{figure}[!ht] 
 \centering
 \includegraphics[width=9cm]{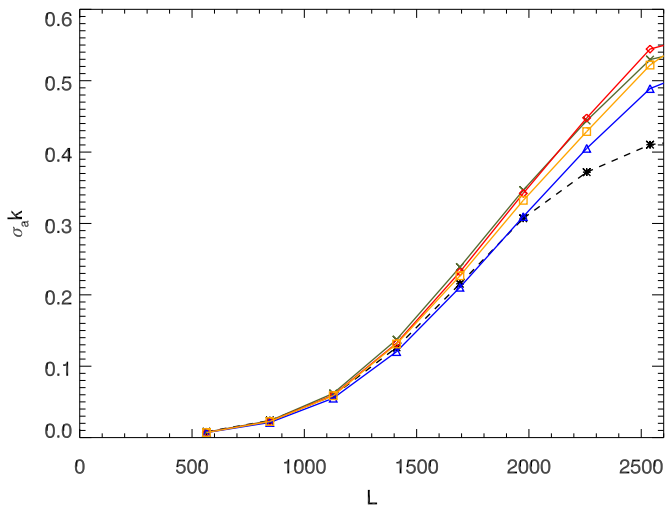}
  \caption{Absorption cross section for pairs of tubes with $D=0.5$ Mm and $\chi=0^o$ (green), $D=0.5$ Mm and $\chi=45^o$ (red), $D=0.5$ Mm and $\chi=90^o$ (orange), $D=1$ Mm and $\chi=0^o$ (blue). The black dashed line corresponds to two non-interacting tubes with a separation of  $D=0.5$ Mm.}
  \label{fig:sigma_a}
\end{figure}

Over the last several years, the dependence of the frequency shifts on the surface magnetic flux has drawn the attention of global helioseismology \citep[\eg][]{Antia+etal2001,Howe+etal2002,Chaplin+etal2007}. These studies are challenged by the insufficient modeling of magnetic regions such as plages. A question of major interest is the quantification of the absorption and phase shift produced by multiple scattering. If these effects could be neglected, the theory of single scattering would provide an accurate description of the scattering produced by group of tubes in the solar atmosphere. This approach was followed by \citet{Jain+etal2009}. However, as shown by most of the simulations analyzed in this work, multiple scattering modifies these measures significantly, and cannot be overlooked. When an incident wave packet reaches a flux tube, part of its energy is absorbed though the excitation of sausage and kink tube waves. These waves propagate upward and downward along the tube and effectively extract energy, so the wave field in the surroundings will be modified. If there is another tube in this wave field, it will produce more absorption, and the variation of the absorption coefficient produced by multiple scattering will depend on how each tube sees the wave field scattered by the companion tube. Figure \ref{fig:sigma_a} shows $\sigma_{\rm a}(L)$ for the different pairs of tubes discussed here. A comparison of the absorption cross section for two non-interacting tubes with those obtained for pairs of interacting tubes reveals that multiple scattering produces an enhacement of the absorption. Because of the scattered wave field produced by one of the tubes, the companion tube receives an acoustic excitation that can exceed what would be expected from the amplitude of the incident wave alone. This excitation depends on the location of the flux tubes, and it is lower when they are perpendicular to the direction of propagation of the incident wave.

Interestingly, taking into account multiple scattering between pairs of tubes provides a much better approximation of the absorption and phase shift produced in bundles of flux tubes with more constituent elements. We found that although one interaction is not a good approximation for the scattering of the $f$ mode caused by a group of tubes, two-interaction scattering may be enough to approximate it, especially at low degree $L$. Thus, truncating the scattering series at two interactions may be a promising approach to simplify future physical models for how plage regions change the wave field.

The scattering produced by a pair of flux tubes depends on the distance between the tubes and their orientation with respect to the direction of propagation of the incident wave. \citet{Hanasoge+Cally2009} showed that the extent of the region of influence of the near field is $\lambda/2$. Most of the separations between tubes considered in this work are well below this limit. Only in the case of the simulation with $d=4$ Mm the wavelengths studied have a $d/\lambda$ larger than 1/2. The effects of multiple scattering in this simulation are negligible.

The distribution of the absorption and phase shift among different azimuthal orders is highly sensitive to the position of the coordinate system used for the Hankel-Fourier decomposition. When the distribution of the flux tubes is symmetric with respect to the line that forms the direction of propagation of the incident wave at the location of the center of the Hankel analysis, multiple scattering keeps the symmetry in the absorption and phase shift between the negative and positive azimuthal orders. Since we have located the coordinate system at the axis of one of the flux tubes, in those cases where $\chi=0^o$ the results are symmetric in azimuthal order. This is due to the fact that the wave field scattered by the first tube in $+m$ and $-m$ are equal, as seen from the location of the second tube. The symmetry is broken when the second tube is located at $\chi\neq 0^o$.

\begin{figure}[!ht] 
 \centering
 \includegraphics[width=9cm]{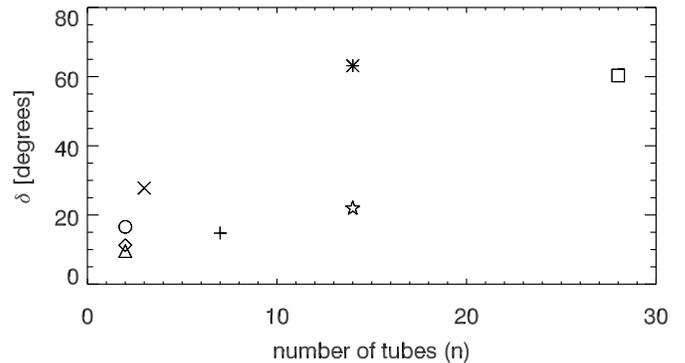}
  \caption{Variation of the phase shift at $L=2539$ and $m=0$ with the number of tubes. Each symbol corresponds to a different case, including two tubes with $d=0.5$ Mm and $\chi =0^o$ (circle), two tubes with $d=1$ Mm and $\chi =0^o$ (diamond), two tubes with $d=4$ Mm and $\chi =0^o$ (triangle), three tubes (cross), seven tubes with $d_{\rm av}=2.14$ Mm (vertical cross), fourteen tubes with $d_{\rm av}=1.07$ Mm (asterisk), fourteen tubes with $d_{\rm av}=2.14$ Mm (star), and twenty-eight tubes with $d_{\rm av}=2.14$ Mm (square). There is a general trend that the phase shift increases with the number of tubes, although the detailed phase shift at a certain $m$ depends on the position of the tubes.}
  \label{fig:phase_ntubes}
\end{figure}

A summary of the phase shifts produced in the axisymmetric mode by bundles of tubes with different number of components is shown in Figure \ref{fig:phase_ntubes}. Since we have performed several simulations with two tubes embedded in the atmosphere, changing the distance between them and the angle $\chi$, we have several data points for this case. We have plotted the phase shift only for those simulations with the tubes aligned with the direction of propagation of the $f-$mode ($\chi =0^o$). We also have several results for the cases with fourteen tubes. The figure shows an increase of the phase shift with the number of tubes. However, note that the phase shift depends not only on the number of tubes, but also on their size, magnetic flux, and relative position, including the separation and angle $\chi$. For example, there is roughly a factor of three difference between the two cases with fourteen tubes, and also ensembles with a large number of constituent elements can produce a smaller phase shift at $m=0$ than other bundles with fewer tubes, just because of the position of the tubes. One would expect that in the Sun waves will not have a preferred angle of incidence to the bundle of flux tubes, and the interaction of the tubes with waves propagating in random directions will result in an azimuthal average.

The simulations of ensembles of randomly distributed tubes are of special interest, since they represent a magnetic structure similar to those expected for plages and other solar features. They show that a group of tubes can absorb quite effectively and also generate phase shifts. Multiple scattering plays a key role in these cases, especially at low $L$ where the parameter $d_{\rm av}/\lambda$ is small and the tubes are strongly coupled by their wavefields. Recently, \citet{Daiffallah2013} found that clusters with a distance between tubes $d/\lambda=0.2$ are better absorbers than more compact clusters. Figure \ref{fig:sigma_ensemble} reveals that the behavior, in general, is more complex than that. Absorption can increase or decrease with the distance depending on $L$ (wavelength). For example, the ratio $d_{\rm av}/\lambda$ for the simulation with $d_{\rm av}=1.07$ at $L=2539$ is the same of the simulation with $d_{\rm av}=3.20$ at $L=846$. However, in the case of the smaller $L$ an increase in the separation between tubes produces a reduction of the total absorption cross section, while high $Ls$ show the opposite response (see inset from Figure \ref{fig:sigma_ensemble}). Note that the methodology of our work is different from \citet{Daiffallah2013}. We have measured the absorption and phase shift by comparing the ingoing and outgoing Hankel components of the total wavefield, whereas \citet{Daiffallah2013} inspects the scattered wavefield, focusing on the temporal evolution at single point. 

An important issue in solar physics consists of checking whether helioseismic measures can help to discern the structure of solar magnetic elements. \citet{Duvall+etal2006} observationally measured the amplitude and phase of monopole and dipole scattering produced in the $f-$mode by thousands of small magnetic elements. \citet{Hanasoge+etal2008} modeled this data assuming that the single scattering approximation captures the nature of the interactions. Recently, \citet{Felipe+etal2012a}, using the same set of data from \citet{Duvall+etal2006}, compared the observed dependence of phase shift with wavenumber and azimuthal order with the results obtained from a single flux tube. Although those magnetic elements are probably composed by ensembles of thin flux tubes, all modeling efforts have avoided the complications associated with multiple scattering. However, multiple scattering leaves a mark in the measured scattering coefficients and, thus, must be taken into account when interpreting observations. 

Retrieving the properties of the magnetic elements (composed by ensembles of flux tubes) presents a serious challenge due to the many parameters involved. Even the most simplistic models should account for the magnetic field and radius of the tubes, and the distances between them. The task is also hindered by the fact that different combinations may yield similar results (\eg, two groups of tubes with different magnetic field strength and number of constituent elements can produce the same absorption, since absorption is enhanced by multiple scattering, or Figure \ref{fig:phase_ntubes} as an example for phase shift). The use of direct observations, in addition to helioseismology, should help by reducing the number of free parameters in the models. Looking at the appropriate very high resolution magnetograms, one could expect to obtain some constraints on the magnetic flux in a magnetic tube or collection of tubes. The development of more sophisticated inversion techniques like, for example, the adjoint method (see \citet{Hanasoge+etal2011} for an application of this method to local helioseismology) must be accomplished. Moreover, a full characterization of multiple scattering is necessary. This work is a step toward this aim, but more modeling is still necessary. The methods presented here could also be used to study wave propagation through models of sunspots, like monolithic or spaghetti sunspots, and also to address additional drivers, which would allow the evaluation of the scattering produced for higher order $p-$modes.

\acknowledgements   This research has been funded by NASA
 through projects NNH09CE43C, NNH09CF68C, NNH09CE41C, and NNH07CD25C. Computer resources were provided by the Extreme Science and Engineering Discovery Environment (XSEDE), which is supported by National Science Foundation grant number OCI-1053575, and NASA's Pleiades supercomputer at Ames Research Center. This work uses Hankel decomposition software developed by D. C. Braun (NWRA) and publicly available at http://www.cora.nwra.com/$\sim$dbraun/hankel/. ACB acknowledges DFG SFB 963 ``Astrophysical Flow Instabilities and Turbulence'' (project A18) aimed at understanding solar and stellar dynamos.

\aareferences

\end{document}